\documentclass[superscriptaddress,twocolumn,prd,showpacs,preprintnumbers,amsmath,amssymb]{revtex4-1}
\usepackage{graphicx}
\usepackage{dcolumn}
\usepackage{bm}
\usepackage{epsfig}
\usepackage{enumerate}
\usepackage{comment}
\usepackage[colorlinks,citecolor=blue,urlcolor=blue,bookmarks=false,hypertexnames=true]{hyperref} 

\def\bk#1#2{\,\langle\,#1|#2 \,\rangle}
\def\bra#1{\,\langle\,#1|}
\def\ket#1{|#1 \,\rangle}

\def\calh{{\cal H}}
\def\calh{{\cal H}}

\def\calm{{\cal M}}

\def\lan{\langle}
\def\ran{\rangle}

\def\aq{\bar{q}}
\def\beq{\begin{equation}}
\def\ee{\end{equation}}
\def\eeq{\end{equation}}

\def\eps{\epsilon}
\def\veps{\varepsilon}

\def\bfig{\begin{figure}}
\def\efig{\end{figure}}
\def\bea{\begin{eqnarray}}
\def\bwt{\begin{widetext}}
\def\ewt{\end{widetext}}
\def\beann{\begin{eqnarray*}}
\def\eea{\end{eqnarray}}
\def\eeann{\end{eqnarray*}}

\def\nn{\nonumber}

\def\3p0{$^{3}P_{0}$}

\def\hs{\hspace{.5cm};\hspace{.5cm}}

\begin{document}

\preprint{APS/123-QED}

\title{Strong decays of strange quarkonia in a corrected \3p0 model}

\author{J. N. de Quadros}
 \affiliation{Instituto de F\'{\i}sica, Universidade Federal do Rio
  Grande do  Sul, Brazil}


\author{D. T.  da Silva}

\author{M. L. L. da Silva}
 
\affiliation{Instituto de F\'{\i}sica e Matem\'atica, 
Universidade Federal de  Pelotas,
 Brazil}


\author{D. Hadjimichef\,}
\email{dimiter.hadjimichef@ufrgs.br; dimihadj@gmail.com}

\affiliation{Instituto de F\'{\i}sica, Universidade Federal do Rio
  Grande do  Sul, Brazil}

\date{\today}

\begin{abstract}

Extensively applied to both light and heavy  meson decay
and standing as  one of the most successful strong decay models is the \3p0 model, in which
$q\bar{q}$ pair production is the dominant mechanism. 
In this paper we evaluate strong decay amplitudes and partial widths of strange $S$ and $D$ state mesons,
namely    $ \phi(1020)$, $\phi (1680) $, $ \phi (2050) $,
$\phi_1 (1850) $, $\phi_2 (1850) $ and $ \phi_3 (1850) $, in the
bound-state corrected \3p0 decay model (C\3p0). The C\3p0 model is obtained 
in the context of the Fock-Tani formalism, which is a mapping technique.

\end{abstract}



\maketitle


\section{Introduction}
\label{intro}

The study of strangeonia should enter a new era with the
advent of the new Hall D photoproduction facility GlueX at
Jefferson Lab \cite{gluex1,gluex2}. 
The main goal of the GlueX experiment is to search for and study hybrid and exotic mesons
which will provide the ideal laboratory for testing QCD in the confinement regime.
Another top goal of GlueX is the exploration of the light meson spectrum, where
interactions of hadrons with a photon beam can be regarded as a superposition of
vector mesons with an important $s\bar{s}$ component. In this sense, studies of
strange final states at GlueX should lead to considerable improvement 
in our knowledge of the $s\bar{s}$ spectrum.

Strange quarkonia are light ($u,d,s$) mesons with at least
one strange quark or antiquark in their dominant $q\bar{q}$ valence
component. These are known as kaonia if the dominant valence basis state is 
$n\bar{s}$ (where $n$ should be understood as $u,d$), antikaonia if $s\bar{n}$, and
strangeonia if $s\bar{s}$.
A principal goal of light meson spectroscopy is the identification of exotica, 
which are resonances that are not dominantly 
$q\bar{q}$ states. These include glueballs, hybrids, and multiquark systems.

In this sense, a great variety of quark-based models are known  that  describe with
reasonable success single-hadron properties. A natural  question
that arises is to what extent a model which gives a good description
of hadron properties is, at the same time, able to describe the
complex hadron-hadron interaction or by the same principles hadron
decay.  In the  direction of clarifying these questions
 is the successful decay model, the \3p0 model, which considers only
OZI-allowed strong-interaction decays.
This model was introduced over thirty years ago by Micu
\cite{micu} and  applied to meson decays in the 1970 by LeYaouanc {\it et al.}
\cite{leyaouanc}. This description is a natural consequence of the
constituent quark model  scenario  of hadronic states.
Since the \3p0 model precedes QCD and has no clear relation to it, 
one might expect that a description of decays in
terms of allowed QCD processes such as OGE might be
more realistic. There is strong experimental evidence that the
$q\bar{q}$ pair created during the decay does have spin one  
as is assumed in  the \3p0   decay model.

T. Barnes {\it  et al.} \cite{barnes1}-\cite{barnes4} have made an
extensive survey of meson states in the
light  of the \3p0 model. Two basic parameters of their formulation are
$\gamma$ (the interaction strength)  and $\beta$ (the wave function's
extension parameter). Although they found the optimum values near
$\gamma= 0.5$ and $\beta=0.4$ GeV, for light $1S$ and $1P$  decays, these
values lead to overestimates of the widths of\,\, higher-$L$ states. In
this perspective a modified $q\bar{q}$ pair-creation interaction, with
 $\gamma= 0.4$ was preferred. The spectrum of meson resonances up 
to 2 GeV is only moderately well determined. For strangeonia the 
they calculated a  set
of  strong decays of a total of 43 resonances into 525 two-body modes, 
with 891 numerically evaluated amplitudes for all energetically allowed open-flavor two-body
decay modes of all $n\bar{s}$ and $s\bar{s} $ strange mesons in the $1S$, 
$2S$, $3S$, $1P$, $2 P$, $1D$ and $1F$ multiplets \cite{barnes_strange}.
 
In the present work, we shall   concentrate on the $\phi$ mesons,
which are the strange $S$ and $D$ states,  predicted in the quark model, 
probable $s\bar{s}$ resonances  expected up to 2.2 GeV.
We employ a mapping technique in order
to obtain an effective interaction for meson decay.
A particular mapping technique
long used in atomic physics \cite{girar1}, the Fock-Tani formalism
(FTf), has been adapted, in previous publications \cite{annals}-\cite{mario}, in order
to describe hadron-hadron scattering interactions with constituent
interchange. Now this technique has been extended in order to
include meson decay \cite{prd08,dimi10}.
Starting with a microscopic $q\bar{q}$ pair-creation interaction,
in lower order, the \3p0 results are reproduced. An additional and
interesting feature appears in higher orders of the formalism: corrections due to
the bound-state nature of the mesons and a natural modification
in the $q\bar{q}$ interaction strength.

In   section \ref{sec:mesons} we review the basic aspects of the formalism.
Section \ref{sec:cap4exame}  is dedicated to obtain an effective
decay Hamiltonian for a $\phi$ meson, where
in subsection \ref{subs1}  the general amplitudes    and decay widths are obtained
with numerical analysis in subsection \ref{subs2}. In section \ref{conc} are
the conclusions.


\section{Meson mapping and the C\3p0 model}
\label{sec:mesons}

\subsection{Review of the Fock-Tani Formalism }
This section reviews the formal aspects of the mapping procedure and
how it is implemented  to quark-antiquark meson states \cite{annals}.
In the Fock-Tani formalism  one starts with the Fock representation of the
system using field operators of elementary constituents which satisfy canonical
(anti) commu\-ta\-tion relations. Composite-particle field operators  are
linear combinations of the elementary-particle operators and do not generally
satisfy canonical (anti) commutation relations. ``Ideal" field
operators acting on an enlarged Fock space are then introduced in close
correspondence with the composite ones.
Next, a given unitary transformation, which transforms the
single composite states into single ideal states, is introduced.
Application of the
unitary operator on the microscopic Hamiltonian, or on
other hermitian operators expressed in terms of the elementary constituent
field operators, gives equivalent operators
which contain the ideal field operators. The effective Hamiltonian
in the new representation
has a clear physical
interpretation in terms of the processes it describes. Since all field
operators in the new representation satisfy canonical (anti)commutation
relations, the standard methods of quantum field theory can then be readily
applied.

The starting point is the definition of single
composite bound states.  We write a single-meson state in terms of a meson
creation operator $M_{\alpha}^{\dagger}$ as
\bea
|\alpha \ran  = M_{\alpha}^{\dagger}|0 \ran ,
\label{1b}
\eea
where $|0 \ran$ is the vacuum state. The meson creation operator $M_{\alpha}^{\dagger}$ is written
in terms of constituent quark and antiquark creation operators
$q^{\dagger}$ and $\aq^{\dagger}$,
\bea
M^{\dagger}_{\alpha}= \Phi_{\alpha}^{\mu \nu}
q_{\mu}^{\dagger} {\aq}_{\nu}^{\dagger} ,
\label{Mop}
\eea
$\Phi_{\alpha}^{\mu \nu}$ is the meson wave function and $q_{\mu}|0
\ran=\aq_{\nu}|0\ran=0$. The index $\alpha$ identifies the meson
quantum numbers of space, spin and isospin. The indices $\mu$ and
$\nu$ denote the spatial, spin, flavor, and color quantum numbers of
the constituent quarks. A sum over repeated indices is implied. 
%
It is convenient to work
with orthonormalized amplitudes,
\bea
\lan\alpha|\beta\ran= \Phi_{\alpha}^{*\mu \nu}
\Phi_{\beta}^{\mu \nu}=\delta_{\alpha \beta}.
\label{norm}
\eea
The quark and antiquark operators satisfy canonical anticommutation relations,
\bea
&&\{q_{\mu}, q^{\dagger}_{\nu}\}=
\{\aq_{\mu},\aq^{\dagger}_{\nu}\}=\delta_{\mu \nu}, \nn\\
&&\{q_{\mu}, q_{\nu}\}=\{\aq_{\mu},\aq_{\nu}\}=
\{q_{\mu}, \aq_{\nu}\}= \{q_{\mu}, \aq_{\nu}^{\dagger}\}=0.
\label{qcom}
\eea
Using these quark anticommutation relations, and the normalization condition
of Eq.~(\ref{norm}), it is easily shown that the meson operators satisfy the
following non-canonical commutation relations
\beq
[M_{\alpha}, M^{\dagger}_{\beta}]=\delta_{\alpha \beta} -
M_{\alpha \beta},\hspace{1.5cm}[M_{\alpha}, M_{\beta}]=0,
\label{M-com}
\eeq
where
\bea
M_{\alpha \beta}= \Phi_{\alpha}^{*{\mu \nu }}
\Phi_{\beta}^{\mu \sigma }\aq^{\dagger}_{\sigma}\aq_{\nu}
+ \Phi_{\alpha}^{*{\mu \nu }}
\Phi_{\beta}^{\rho \nu}q^{\dagger}_{\rho}q_{\mu}.
\label{delta}
\eea
A transformation is defined such that  a single-meson state
$|\alpha \ran$ is redescribed by an (``ideal") elementary-meson state by
\bea
|\alpha \ran\longrightarrow U^{-1}|\alpha\rangle =
m^{\dagger}_{\alpha}|0\rangle,
\label{single_mes}
\eea
where $m^{\dagger}_{\alpha}$ an ideal meson creation operator. The ideal
meson operators $m^{\dagger}_{\alpha}$ and $m_{\alpha}$ satisfy,
by definition, canonical commutation relations
\beq
[m_{\alpha}, m^{\dagger}_{\beta}]=\delta_{\alpha \beta} ,
\hspace{1.5cm}[m_{\alpha}, m_{\beta}]=0.
\label{mcom}
\eeq
The state $|0\rangle$ is the vacuum of both $q$ and $m$ degrees of freedom in the
new representation.
In addition, in the new representation the quark and antiquark operators
$q^{\dagger}$, $q$, $\aq^{\dagger}$ and $\aq$ are kinematically independent of
the $m^{\dagger}_{\alpha}$ and $m_{\alpha}$
\bea
[q_{\mu},m_{\alpha}]=[q_{\mu},m^{\dagger}_{\alpha}]=[\aq_{\mu},m_{\alpha}]=
[\aq_{\mu},m^{\dagger}_{\alpha}]=0 .
\label{indep_mes}
\eea
The
unitary operator $U$ of the transformation is 
\bea 
U(t)=\exp\left[t\, F\right] , 
\label{u} 
\eea 
where $F$ is the generator of the
transformation and $t$ a parameter which is set to $-\pi/2$ to
implement the mapping. The generator $F$ of the transformation is
\bea 
F= m^{\dag}_{\alpha}\,\tilde{M}_{\alpha}-
\tilde{M}^{\dag}_{\alpha} m_{\alpha} 
\label{f-generator} 
\eea 
where
\bea 
\tilde{M}_{\alpha}=\sum_{i=0}^{n}\tilde{M}^{(i)}_{\alpha},
\label{mes_gen} 
\eea 
with 
\bea
&&[\tilde{M}_{\alpha},\tilde{M}^{\dagger}_{\beta}] =
\delta_{\alpha\beta}
\hspace{.5cm} + \hspace{.5cm}{\cal O} (\Phi^{n+1}),\nn\\
&&[\tilde{M}_{\alpha}, \tilde{M}_{\beta}]= [
\tilde{M}^{\dagger}_{\alpha},  \tilde{M}^{\dagger}_{\beta}]=0.
\label{comO} 
\eea 
It is easy to see from (\ref{f-generator}) that
$F^{\dag}=-F$ which ensures that $U$ is unitary. The index $i$ in
(\ref{mes_gen}) represents the order of the expansion in powers of
the wave function $\Phi$. The $\tilde{M}_{\alpha}$ operator is
determined up to a specific order $n$ consistent with (\ref{comO}).

The next step is to obtain the transformed operators in the new
representation. The basic operators of the model are expressed in
terms of the quark operators. Therefore, in order to obtain the
operators  in the new representation, one writes
\bea
q(t)=U^{-1}\, q \,U,\hspace{1.0cm}{\aq}(t)=U^{-1}\, {\aq}\, U .
\eea

Once a microscopic interaction Hamiltonian $H_I$ is defined, at the quark
level, a new transformed Hamiltonian can be obtained. This effective
interaction,  the {\sl Fock-Tani Hamiltonian} ($\calh_{\rm FT}$),  is
obtained   by the application of the unitary operator
$U$ on the microscopic Hamiltonian $H_I$, {\it i.e.}, $\calh_{\rm FT}=U^{-1}\,H_I\,U$.
The transformed Hamiltonian   describes all possible
processes involving mesons and quarks.
The general structure of $\calh_{\rm FT}$  is of  the following form
\bea
\calh_{\rm FT}=\calh_{\rm q} + \calh_{\rm m} + \calh_{\rm m q} ,
\label{separation}
\eea
where the first term involves only quark operators, the second one involves
only ideal meson operators, and $\calh_{\rm m q}$ involves quark and
meson operators.
In $\calh_{\rm FT}$ there are higher order terms that provide bound-state corrections
to the lower order ones. The basic quantity for these corrections is the {\it bound-state kernel}
$\Delta$ defined as
\begin{eqnarray}
\Delta(\rho\tau;\lambda\nu)
=\Phi^{\rho\tau}_{\xi}\Phi^{\ast\lambda\nu}_{\xi}.
\label{kernel}
\end{eqnarray}
The physical meaning of the $\Delta$ kernel becomes evident, in the
sense that it   modifies  the quark-antiquark interaction strength \cite{annals,prd08}.
The following two examples can   clarify the physical interpretation.

(1) \underline{\it First   example}: consider that the  starting point is a
two-body microscopic quark-antiquark Hamiltonian of the form 
\bea
H_{2q} &=& T\left(\mu\right)q^{\dagger}_{\mu} q_{\mu} + T\left(\nu\right)
\aq_{\nu}^{\dagger}\aq_{\nu} 
+ V_{q\aq}(\mu\nu;\sigma\rho)q^{\dagger}_{\mu}\aq^{\dagger}_{\nu}
\aq_{\rho}q_{\sigma} 
\nn\\
&&
\!\!\!\!
+ \frac{1}{2} V_{qq}(\mu\nu;\sigma\rho)
q^{\dagger}_{\mu}q^{\dagger}_{\nu}q_{\rho}q_{\sigma}
+\frac{1}{2}V_{\aq\aq}(\mu\nu;\sigma\rho)\aq^{\dagger}_{\mu}
\aq^{\dagger}_{\nu}\aq_{\rho}\aq_{\sigma}.
\nn\\
\label{qHamilt}
\eea
The transformation $\calh_{\rm FT}=U^{-1}\,H_{2q}\, U$ is implemented again by  
transforming each quark and antiquark 
operator in Eq.~(\ref{qHamilt}), where a  similar structure to
Eq. (\ref{separation}) is obtained.
In free space, the wave function $\Phi$ of Eq.~(\ref{Mop}) satisfies the following 
equation
\beq
H(\mu\nu; \sigma\rho)\Phi_{\alpha}^{\sigma\rho}=\epsilon_{[\alpha]} 
\Phi_{[\alpha]}^{\mu \nu},
\label{Schro}
\eeq
where $H(\mu\nu; \sigma\rho)$ is the Hamiltonian matrix
\bea
H(\mu\nu; \sigma\rho)&=&\delta_{\mu[ \sigma]} \delta_{\nu[\rho]} 
\left[T([\sigma]) + T([\rho])\right]
\nn\\
&&+ 
V_{q\aq}(\mu\nu; \sigma\rho) ,
\label{Schroperat}
\eea
$\epsilon_{[\alpha]}$ is the total energy of the meson. There is no sum 
over repeated indices inside square brackets.

The  effective quark Hamiltonian $\calh_{\rm 2q}$  has an identical
structure to the  microscopic quark 
Hamiltonian, Eq.~(\ref{qHamilt}), except for inclusion of a term corresponding to a
modification in the quark-antiquark interaction as follows
\bea
\calh_{\rm 2q}={H}_{2q}+\bar{H}_{q\bar{q}}\,,
\label{modqaq1}
\eea
with
\bea
\bar{H}_{q\bar{q}}=\left(-H\,\Delta - \Delta\,H     +  \Delta\,H\,\Delta\,\right)\,
q^{\dagger}_{\mu}\aq^{\dagger}_{\nu}
\aq_{\rho}q_{\sigma} 
\label{modqaq}
\eea
%
%
where the contractions are 
$H\,\Delta\equiv H(\mu \nu;\tau\xi)\,\Delta(\tau\xi; \sigma \rho)$ and
$\Delta H\,\Delta\equiv \Delta(\mu \nu;\zeta\eta) H(\zeta\eta;\tau\xi)\,\Delta(\tau\xi; \sigma \rho)$.
An important  property of the bound-state kernel is 
\beq
\Delta(\mu \nu; \sigma\rho)\Phi^{\sigma\rho}_{\alpha}=\Phi^{\mu\nu}_{\alpha},
\label{propDelta}
\eeq
which follows from the wave function's orthonormalization, Eq.~(\ref{norm}).
In the case that $\Phi$ is a solution of
Eq.~(\ref{Schro}), equation (\ref{modqaq1}), reduces to
\bea
\calh_{\rm 2q}=H_{2q}-\eps_{\beta}\,N_{\beta}
\label{modqaq2}
\eea
where $N_{\beta}=M^{\dag}_\beta\,M_\beta\,$ is the number operator and the
following property holds:
$N_{\beta}|\alpha\rangle=|\beta\rangle$\,.

The spectrum of the modified quark 
Hamiltonian, $\calh_{\rm 2q}$, is positive semi-definite and hence has
no bound-states~\cite{girar1}. To show this, consider an arbitrary state
 $| \alpha\rangle$ formed from a pair quark and antiquark: 
\beq
|\alpha\rangle  = \Psi_\alpha^{\mu \nu}q_{\mu}^{\dag}{\aq}_{\nu}^{\dag}|0\rangle\;.
\label{II.84}
\eeq
The action of the Hamiltonian (\ref{modqaq2}) on this state results in
\bea 
\calh_{\rm 2q}| \alpha\rangle&=& 
\left(\frac{}{} H_{2q}-\eps_{\beta}\,N_{\beta}\,\right)| \alpha\rangle
\nn\\
&=&  H_{2q}| \alpha\rangle-\eps_{\beta}\,\Phi_\beta^{\ast\,\mu \nu}\Psi_\alpha^{\mu \nu} |\beta\rangle
\label{II.85a}
\eea
If $ |\alpha\rangle$, is one of the   bound eigenstates of the microscopic  Hamiltonian
then
\bea 
 \calh_{\rm 2q}|\alpha\rangle= 
\left(\frac{}{} \eps_{[\alpha]}-\eps_{[\alpha]}\,\right)|[\alpha]\rangle=0\,.
\label{II.85}
\eea
 On the other hand, if  $\Psi_\alpha^{\mu \nu}$ is orthogonal to all bound states $\Phi_\alpha^{\mu \nu}$ then (\ref{II.85a}) 
reduces to
\bea 
\calh_{\rm 2q}| \alpha\rangle&=&  H_{2q}| \alpha\rangle\,.
\label{II.87}
\eea
Let $ \psi_{i\,\alpha}$ be the continuum (unbound, positive energy)
eigenstates of $H_{2q}$, with energies $\veps_{i\,\alpha} \geq 0$\,.
One can expand any $ \Psi_\alpha^{\mu \nu}$ in the form 
\bea
\Psi_{ \alpha}^{\mu \nu} =
\sum_\kappa \,c_{\kappa}\Phi_{\kappa\alpha}^{\mu \nu}  +
\sum_i \,c_{i} \psi_{i\,\alpha}^{\mu \nu}\;,
\label{II.89}
\eea
%
%
%
%
where $(\Phi_\kappa,\psi_i )=0$\,.
Then by   (\ref{II.85}) and  (\ref{II.87})
\bea
\calh_{\rm 2q}|\alpha\rangle=  
\sum_{i} \veps_{i\,\alpha}\,c_{i}\,\psi_{i\,\alpha}^{\mu \nu}\,|\mu \nu\rangle\;.
\label{II.90}
\eea
Therefore,
\beq
\langle  \alpha|\calh_{\rm 2q}| \alpha\rangle= \sum_{i}
\veps_{i\,\alpha}|c_{i}|^{2}\;,
\label{II.91}
\eeq
where it is evident that $ \calh_{\rm 2q}$ is semi-definite positive and therefore does not have
quark-antiquark bound states.

(2) \underline{\it Second   example}:
consider in the ideal meson sector ${\cal H}_{\rm \bf m}$
of equation (\ref{separation}), many  approaches similar to the Fock-Tani formalism \cite{annals}  
have obtained, for example, the meson-meson scattering interaction in the Born
approximation: Resonating Group Method (RGM) \cite{oka}, Quark Born Diagram Formalism (QBDF) \cite{qbd},

\beq
H_{mm}= T_{mm} +V_{mm},
\eeq
where $T_{mm}$ is  the kinetic term and $V_{mm}$  is the meson-meson interaction potential with constituent interchange.
This potential is given by
\bea
V_{mm}=V_{mm}^{dir}+V_{mm}^{exc}+V_{mm}^{int} \,,
\eea
where $V_{mm}^{dir}$ is the direct 
potential (no quark interchange), $V_{mm}^{exc}$ the quark exchange term and $V_{mm}^{int}$ 
the intra-exchange term.  As shown in Ref. \cite{annals} and \cite{sergio}, if one extends the Fock-Tani calculation to higher orders
a new meson-meson Hamiltonian is obtained
\bea
\bar{H}_{mm}=H_{mm}+\delta H_{mm}
\label{h-mm}
\eea
where $\delta H_{mm} $ is the bound-state correction Hamiltonian,
\bea
\delta H_{mm}
&=&
\frac{1}{2}\Phi_{\alpha}^{*\mu\nu}\Phi_{\beta}^{*\rho\sigma}
H(\mu\nu;\lambda\tau)\Delta(\lambda\tau;\mu^{\prime}\sigma^{\prime})
\Phi_{\delta}^{\mu^{\prime}\sigma}\Phi_{\gamma}^{\rho\sigma^{\prime}}
\nn\\
&+&\frac{1}{2}\Phi_{\alpha}^{*\rho\sigma}
\Phi_{\beta}^{*\mu\nu}H(\mu\nu;\lambda\tau)\Delta(\lambda\tau;\mu^{\prime}
\sigma^{\prime})\Phi_{\delta}^{\rho\sigma^{\prime}}
\Phi_{\gamma}^{\mu^{\prime}\sigma}\nn\\  
&+&\frac{1}{2}\Phi_{\alpha}^{*\mu\sigma}
\Phi_{\beta}^{*\rho\nu}\Delta(\mu\nu;\lambda\tau)H(\lambda\tau;\mu^{\prime}
\nu^{\prime})\Phi_{\delta}^{\mu^{\prime}\nu^{\prime}}
\Phi_{\gamma}^{\rho\sigma}\nn\\
&+&\frac{1}{2}\Phi_{\alpha}^{*\rho\nu}
\Phi_{\beta}^{*\mu\sigma}\Delta(\mu\nu;\lambda\tau)
H(\lambda\tau\mu^{\prime}\nu^{\prime})\Phi_{\delta}^{\rho\sigma}
\Phi_{\gamma}^{\mu^{\prime}\nu^{\prime}}\,.
\nn\\
\eea
If the wave function  $\Phi$ is chosen to be an eigenstate of the microscopic quark 
Hamiltonian, then the intra-exchange term $V_{mm}^{int}$ is cancelled 
exactly:
\bea
V_{mm}^{int}+\delta H_{mm}=0.
\label{intra-cancel}
\eea
In summary, these examples reveal an important and
common feature of  these corrections to the leading order: they modify
the microscopic potential in the presence of bound-states.

\subsection{Meson decay in the Fock-Tani Formalism }

In the present calculation, the microscopic interaction Hamiltonian
is a  pair creation Hamiltonian $H_{q\aq}$  defined as
\bea H_{q\aq}=V_{\mu\nu}\,
q^{\dag}_{\mu}\aq^{\dag}_{\nu} \,,
\label{h_3p0}
\eea
where in (\ref{h_3p0})
 a sum (integration) is again  implied  over repeated indexes \cite{prd08}.
%
The pair creation potential $V_{\mu\nu}$ is given by
\bea
V_{\mu\nu}\equiv
 g\,
\delta_{c_{\mu}c_{\nu}}\delta_{f_{\mu}f_{\nu}}
\delta(\vec{p}_{\mu}+\vec{p}_{\nu})\,
\bar{u}_{s_{\mu}} (\vec{p}_{\mu}) \, 
v_{s_{\nu} }(\vec{p}_{\nu}) ,
\label{vmn}
\eea
with $g=2\,m_{q}\, \gamma\,$, where $\gamma$ is the pair production strength
and the indexes $c_\mu$, $f_\mu$, $s_\mu$ are of color, flavor and spin. 
The pair production is obtained from the  non-relativistic limit
 of $H_{q\aq}$ involving Dirac quark fields \cite{barnes1}.
Applying the Fock-Tani transformation to $H_{q\aq}$ one obtains the effective
Hamiltonian that describes a decay process.
In the FTf perspective a new
 aspect is introduced  to meson decay: bound-state corrections.
The lowest order correction is one that involves only one
bound-state kernel $\Delta$.
The bound-state corrected, C\3p0 Hamiltonian, is
\bea
\!\!\!\!\!\!\!\!\!\!\!\!\!\!\!\!
H^{\rm C3P0}
&=& -\Phi^{\ast\rho\xi}_{\alpha} \Phi^{\ast\lambda\tau}_{\beta}
\Phi^{\omega\sigma}_{\gamma}\, V^{\rm C3P0}\, m^{\dag}_{\alpha}
m^{\dag}_{\beta} m_{\gamma},
\label{c3p0}
\eea
where $V^{\rm C3P0} $
is a condensed notation for 
\bea
V^{\rm C3P0}&=&
\left(\,\bar{\delta}
+\bar{\Delta}\,\right)\,V_{\mu\nu}\,,
\label{vc3p0}
\eea
where
\bea
\bar{\delta}&=&\delta_{\mu\lambda}
\delta_{\nu\xi}
\delta_{\omega\rho}
\delta_{\sigma\tau}
\nn\\
\bar{\Delta}&=&
\frac{1}{4}\left[
\delta_{\sigma\xi}\,\delta_{\lambda\mu}\,\,
\Delta(\rho\tau;\omega\nu)
\frac{}{}
+
\delta_{\xi\nu}\,\delta_{\lambda\omega}\,\,
\Delta(\rho\tau;\mu\sigma)
\right.
\nn\\
&&
\left. \frac{}{}
-2\delta_{\sigma\xi}\,\delta_{\lambda\omega}\,\,
\Delta(\rho\tau;\mu\nu)
\right]\,.
\label{deltas}
\eea
%
%
%
The first term of (\ref{vc3p0}), involving 
$\bar{\delta}$ is the usual  \3p0 decay potential. The following $\bar{\Delta}$ term,   containing three  $\Delta$'s, is the 
bound-state correction to the potential.
In the ideal meson space the initial and final states involve only ideal
meson operators $|i\rangle=m^{\dag}_{\gamma}|0\rangle$ and
$|f\rangle=m^{\dag}_{\alpha}m^{\dag}_{\beta} |0\rangle $.
The C\3p0 amplitude is obtained by the following matrix element,
\bea
\hspace{-1cm}
\langle f | H^{\rm C3P0} | i \rangle &=&
\delta(\vec{P}_\gamma-\vec{P}_\alpha-\vec{P}_\beta)\, h_{fi}^{\rm C3P0}
\label{ideal-matrix}
\eea
The $h_{fi}^{\rm C3P0}$ decay amplitude is combined with
relativistic phase space, resulting in the differential decay rate
\bea
\frac{d\Gamma_{\gamma\to \alpha\beta}}{d\Omega}=2\pi\,P\,
\frac{E_\alpha\,E_\beta}{M_\gamma}|h_{fi}^{\rm C3P0}|^2
\label{dif-gamma}
\eea
which, after integration in the solid angle $\Omega$, a  usual choice for the meson
momenta is made: $\vec{P}_\gamma=0$
($P=|\vec{P}_\alpha|=|\vec{P}_\beta|$).


\section{ $\phi$ Meson Decay}
\label{sec:cap4exame}

The previous section has outlined the 
essential aspects of the $C^{3}P_{0}$ model and how it is 
obtained from the Fock-Tani formalism, where the decay
Hamiltonian $H^{C3P0}$ was deducted. In this section the phenomenological Hamiltonian $H^{C3P0}$
will be used   in order to  evaluate the $n^{\,3}S_1$ decays   $ \phi(1020)$, $\phi (1680) $, $ \phi (2050) $
with $n=1,2,3$ 
and the $1^{3}D_J$ decays $\phi_1 (1850) $, $\phi_2 (1850) $ and $ \phi_3 (1850) $ mesons.

\subsection{Amplitudes and decay widths}
\label{subs1}

\indent
 
In the following decay channels, that shall be studied, some
have been observed, with no available data, while others  
are only theoretical \cite{pdg,barnes_strange}: 
\begin{enumerate}[(a)]
\item $\phi(1020)\to KK$;

\item $\phi(1680)\to KK$, $KK^*$, $\eta \phi$;

\item $^{\ddag\,}\phi(2050)\to KK$, $KK^*$, $K^*K^*$, $KK_1(1270)$, $
KK_1(1400)$, $KK_0^*(1430)$, $KK_2^*(1430)$, $KK^*(1410)$,
$KK(1460)$, $\eta \phi$, $\eta^\prime \phi$, $\eta h_1(1380)$;

\item $^{\ddag\,}\phi_1(1850)\to K^*K^*$; 

\item $^{\ddag\,}\phi_2(1850)\to KK$, $ KK^*$, $K^*K^*$, $\eta \phi$; 

\item $\phi_3(1850)\to KK$, $KK^*$, $K^*K^*$, $KK_1(1270)$, $ \eta \phi$,

\end{enumerate}
where the $\ddag\,$ symbol indicates an unobserved state, $\phi_1 (1850)$ is sometimes referred  only as 
$\phi(1850)$ and $K^\ast$ is actually $K^\ast(892)$ from the Particle Data Group \cite{pdg}.
In the calculation of the decay amplitudes, the matrix element  is given by (\ref{ideal-matrix})
where   the decay Hamiltonian (\ref{c3p0}) can be split into two parts: $H^{\rm C3P0}=H_m+\delta H_m$.
The the matrix element of  first term, containing $H_m$,
the  term without the bound-state correction, 
is given by
\begin{eqnarray}
\left\langle f\left|H_{m}\right|i\right\rangle =-d_{1}-d_{2}
\label{4.1-4a}
\end{eqnarray}
where
\begin{eqnarray}
d_{1}&=&\Phi_{\alpha}^{\ast\rho\nu}\Phi_{\beta}^{\ast\mu\lambda}
\Phi_{\gamma}^{\rho\lambda}V_{\mu\nu}
\nonumber \\
d_{2}&=&\Phi_{\alpha}^{\ast\mu\lambda}
\Phi_{\beta}^{\ast\rho\nu}\Phi_{\gamma}^{\rho\lambda} V_{\mu\nu}\,.
\label{4.1-4b}
\end{eqnarray}
The matrix element of the bound-state correction $\delta H_m$, is written as
\begin{eqnarray}
\left\langle f\left|\delta H_m\right|i\right\rangle   =
-\sum_{k=1}^{3}\,\sum_{j=1}^{2} \,d_{j}^{\,k}
\label{4.3-2}
\end{eqnarray}
where we introduce the following notation
\begin{eqnarray}
d_{1}^{1} & = & \frac{1}{4}\Phi_{\alpha}^{\ast\rho\sigma}
\Phi_{\beta}^{\ast\mu\tau} \Delta(\rho\tau;\lambda\nu)
\Phi_{\gamma}^{\lambda\sigma}V_{\mu\nu}\nonumber \\
d_{2}^{1} & = & \frac{1}{4}\Phi_{\alpha}^{\ast\mu\tau}
\Phi_{\beta}^{\ast\rho\sigma} \Delta(\rho\tau;\lambda\nu)
\Phi_{\gamma}^{\lambda\sigma}V_{\mu\nu}\nn\\
d_{1}^{2} & = & -\frac{1}{2}\, \Phi_{\alpha}^{\ast\rho\sigma}
\Phi_{\beta}^{\ast\lambda\tau} \Delta\left(\rho\tau;\mu\nu\right)
\Phi_{\gamma}^{\lambda\sigma}V_{\mu\nu}\nn\\
d_{2}^{2} & = &-\frac{1}{2}\, \Phi_{\alpha}^{\ast\lambda\tau}
\Phi_{\beta}^{\ast\rho\sigma} \Delta\left(\rho\tau;\mu\nu\right)
\Phi_{\gamma}^{\lambda\sigma}V_{\mu\nu} \nn\\
d_{1}^{3} & = & \frac{1}{4}\Phi_{\alpha}^{\ast\rho\nu}
\Phi_{\beta}^{\ast\lambda\tau} \Delta(\rho\tau;\mu\sigma)
\Phi_{\gamma}^{\lambda\sigma}V_{\mu\nu}\nn\\
d_{2}^{3} & = &\frac{1}{4}\Phi_{\alpha}^{\ast\lambda\tau}
\Phi_{\beta}^{\ast\rho\nu} \Delta(\rho\tau;\mu\sigma)
\Phi_{\gamma}^{\lambda\sigma}V_{\mu\nu}\,. 
\label{4.3-5c}
\end{eqnarray}
In $d_{1(2)}^k$, the index $k=1,2,3$ represents the first, second and third
term of the correction, respectively. 
As can be seen in equations (\ref{4.1-4a})-(\ref{4.3-5c}) the matrix elements
depend   directly  on the  the wave functions $\Phi_{\alpha}^{\mu\nu}$ and
the potential $V_{\mu\nu}$. Considering as the fundamental  degrees of freedom
color $C$ , flavor $f$, spin $\chi$ and space $\Phi$, the mesons wave function can be written as
product
\bea \Phi_{\alpha}^{\mu\nu}=
C^{c_{\mu}c_{\nu}} f_{f_{\alpha} }^{f_{\mu}f_{\nu}} \chi_{
S_{\alpha} }^{ s_{\mu}s_{\nu} } \Phi_{nl
}(\vec{P}_{\alpha}-\vec{p}_{\mu}-\vec{p}_{\nu}) \;,
\label{funcdomeson} 
\eea
allowing to calculate color, flavor and spin-space separately.
Details of (\ref{funcdomeson}) can be found in the appendix \ref{funcao-onda}.
This factorization of the wave function implies that equations (\ref{4.1-4b}) and (\ref{4.3-5c}) 
can also be put in a direct product form of color, flavor and spin-space:
\bea
d_1&=&d_1^c\,d_1^f\,d_1^{s-s}
\hs
d_2=d_2^c\,d_2^f\,d_2^{s-s} 
\label{d1-2}
\eea
and
\bea
d_{1}^{1} & = & \frac{1}{4}d_1^{1c}\,d_1^{1f}\,d_1^{1s-s}
\hs
d_{2}^{1}  =  \frac{1}{4}d_2^{1c}\,d_2^{1f}\,d_2^{1s-s}
\nn\\
d_{1}^{2} & = & -\frac{1}{2}\, d_1^{2c}\,d_1^{2f}\,d_1^{2s-s}
\hs
d_{2}^{2}  = -\frac{1}{2}\, d_2^{2c}\,d_2^{2f}\,d_2^{2s-s}
\nn\\
d_{1}^{3} & = & \frac{1}{4}d_1^{3c}\,d_1^{3f}\,d_1^{3s-s}
\hs
d_{2}^{3}  = \frac{1}{4}d_2^{3c}\,d_2^{3f}\,d_2^{3s-s}\,.
\label{d1-2c}
\eea
It is essential to note that the
 bound-state kernel definition, Eq. (\ref{kernel}),
has an implicit contraction in the $\xi$ index, which physically implies 
a {\it sum over all  species} condition. In practice, this means that
one should sum over intermediate meson bound-states.
Any of the respective meson multiplet members can be considered in this sum.
In our calculations, due to the symmetry of
problem, the only possible states will have the following $n\;^{2S+1}L_J$ 
and isospin quantum numbers: $\left|1\;^1S_0\right\rangle$ and
 $I=0$  (type $\eta$, $\eta^\prime$) or
$\left|1\;^3S_1\right\rangle$ and $I=0$ (type $\phi$, $\omega$).
The bound-state kernel $\Delta(\mu\nu;\rho\sigma)$ will then be a sum over
$\eta$, $\eta^\prime$, $\phi$  and $\omega$ intermediate  states,
 which can be  written explicitly as
\bea 
\Delta(\mu\nu;\rho\sigma)&=&\Delta_\eta(\mu\nu;\rho\sigma)
+\Delta_{\eta^\prime}(\mu\nu;\rho\sigma)\nn\\
&+& \Delta_\phi(\mu\nu;\rho\sigma)+\Delta_\omega(\mu\nu;\rho\sigma).
\label{kernel-cor} 
\eea
%
%
The color amplitude factors of  (\ref{d1-2}) and (\ref{d1-2c})  can be calculated
directly  with the definition (\ref{color}) and the color part of (\ref{vmn}), resulting in 
\begin{eqnarray}
d_{1}^{c} & = & d_{2}^{c}=\frac{1}{\sqrt{3}}\,.
\label{4.1-25}
\end{eqnarray}
Proceeding similarly,  for the bound-state correction one has
\begin{eqnarray}
d_{1}^{1c}&=&d_{2}^{1c}
= \frac{d_{1}^{\,2c}}{3 }= \frac{d_{2}^{\,2c}}{3 }
=d_{1}^{3c}=d_{2}^{3c} 
= \frac{1}{3\sqrt{3}}\,.
\label{4.1-25cor}
\end{eqnarray}
This result is independent of which meson is  involved, Eqs.
(\ref{4.1-25}) and (\ref{4.1-25cor}) are valid for all decay processes.
The flavor factor will be evaluated in the next section for each case. 
General spin-space amplitude factors can be obtained from 
the matrix element (\ref{d1-2}) and (\ref{d1-2c}), apart from
a global  momentum conservation  $\delta$.
The contribution
without  the bound-state correction is
\begin{eqnarray}
d_{1}^{s-s} & = &  -2\,a_{ij}\,\gamma \int d^{3}
{K}\,\chi_{i}^{\ast}\left(\vec{\sigma}
\cdot\vec{K}\right)\chi_{j}^{c}\, \phi^\ast\left(2\vec{ {K}}+
{\vec{P}}\right)\nn\\
&&\times\phi^\ast\left(2\vec{ {K}}+ { {\vec{P}}}\right)
\phi\left(2\vec{ {K}}+2 {\vec{P}}\right)
\label{4.1-44}
\end{eqnarray}
and the three terms with the bound-state correction are
\begin{eqnarray}
d_{1}^{1s-s} & = & -2\,a_{1_{ij}}\,\gamma \int d^{3} {K}\, d^{3}
{q}\,\chi_{i}^{\ast}\left(\vec{\sigma}
\cdot\vec{K}\right)\chi_{j}^{c}\, \phi^\ast\left(2\vec{ {q}}+\vec{
{P}}\right)\nonumber \\
 &&\times \phi^\ast\left(2\vec{ {K}}+\vec{ {P}}\right)\left[
\phi\left(\vec{ {q}}+\vec{ {K}}+2\vec{ {P}}\right)
\phi^\ast\left(\vec{ {q}}+\vec{ {K}}\right)\right]_\xi
\nonumber \\
 &&\times \phi\left(2\vec{
{q}}\right) 
\nn\\
d_{1}^{2s-s} & = & -2\,a_{2_{ij}}\,\gamma \int d^{3} {K}\, d^{3}
{q}\,\chi_{i}^{\ast}\left(\vec{\sigma}
\cdot\vec{K}\right)\chi_{j}^{c}
 \phi^\ast\left(2\vec{ {q}}+\vec{ {P}}\right)
\nonumber \\
 &&\times\phi^\ast\left(2\vec{ {q}}+\vec{ {P}}\right)\left[ \phi\left(2\vec{
{q}}+2\vec{ {P}}\right) \phi^\ast\left(2\vec{ {K}}\right)\right]_\xi
\phi\left(2\vec{ {q}}\right)
\nn
\\ 
d_{1}^{\,3s-s} & = & -2\,a_{3_{ij}}\,\gamma \int d^{3} {K}\, d^{3}
{q}\,\chi_{i}^{\ast}\left(\vec{\sigma}
\cdot\vec{K}\right)\chi_{j}^{c} \phi^\ast\left(2\vec{ {q}}-\vec{
{P}}\right)\nonumber \\
 &&\times \phi^\ast\left(2\vec{ {K}}-\vec{ {P}}\right)
\left[\phi\left(\vec{ {q}}+\vec{ {K}}-2\vec{ {P}}\right)
\phi^\ast\left(\vec{ {q}}+\vec{ {K}}\right)\right]_\xi 
\nonumber \\
 &&\times\phi\left(2\vec{
{q}}\right), 
\label{4.3-15c}
\end{eqnarray}
where $a_{ij}=\chi_{\alpha}^{ s_{\rho}s_{\nu}} \chi_{\beta}^{
s_{\mu}s_{\lambda}}\chi_{\gamma}^{ s_{\rho}s_{\lambda}}$ is a
number resulting from the product of the meson's spin wave functions 
involved in the decay. The coefficients
$a_{1_{ij}}$, $a_{2_{ij}}$ and $a_{3_{ij}}$ are obtained in a similar
form and represent  the first, second and third bound-state  correction term, respectively.
Note that the wave functions in-between brackets in (\ref{4.3-15c})  are related to the 
bound-state kernel part and therefore it is assumed 
there is an implicit sum over species 
with  $\xi$ assuming the $\eta,\eta^\prime,\phi,\omega$ quantum numbers.
The $d_2^{s-s}$ and $d_2^{is-s}$ amplitude factors  are obtained   simply by the changing
$\vec{P}\to-\vec{P}$ in (\ref{4.1-44})-(\ref{4.3-15c}).



\subsection{Numerical results }  
\label{subs2}

\subsubsection{General aspects}
In this section, we present the numerical results for the $\phi_J(M)$ decay widths.
The  amplitudes can be written in a general form
\bea 
h_{fi}^{\rm C3P0} & = &
\frac{\gamma}{\pi^{1/4}} \,\calm_{fi} 
\label{hfi} 
\eea 
where $\calm_{fi}$ appears  in  appendix \ref{amplitudes}. These amplitudes are inserted in
(\ref{dif-gamma}) and integrated over the solid angle $\Omega$.
In order to calculate $\calm_{fi}$, the wave function must
be determined, knowing the  spin and
space quantum numbers to be used, which are  listed in table
\ref{notacao-espectroscopica}. 
The spatial wave functions are considered to be Gaussians characterized by 
$\beta$ parameter, which is the Gaussian's width.
Each decay particle  has its own $\beta$.
For example,  $\phi(1020)$ has the width
$\beta_{\phi}$,  $\phi(1680)$ has $\beta_{\phi_{1680}}$ and so
on. The mesons which are part of the bound-state kernel  also have their own widths, and
are distinguished from others by the notation $\beta_{\eta_\Delta}$,
$\beta_{\eta_\Delta^\prime}$, $\beta_{\phi_\Delta}$ and
$\beta_{\omega_\Delta}$.
\begin{table}[!h]
\begin{center}
\begin{tabular}{ c|c}
  $n\;^{2S+1}L_J$ & meson \\
  \hline
  $1\;^1S_0$ & $\eta, \eta^\prime,\, K$ \\
  $1\;^3S_1$ & $\phi(1020),\, K^*$ \\
  $1\;^1P_1$ & $h_1(1380)$ \\
  $1\;^3P_0$ & $ K_0^*(1430)$ \\
  $1\;^3P_2$ & $K_2^*(1430)$ \\
  $1\;^3D_1$ & $\phi_1(1850)^\ddag$ \\
  $1\;^3D_2$ & $\phi_2(1850)^\ddag$ \\
  $1\;^3D_3$ & $\phi_3(1850)$ \\
  $2\;^1S_0$ & $K(1460)$ \\
  $2\;^3S_1$ &$\phi(1680),\, K^*(1410)$ \\
  $3\;^3S_1$ &$\phi(2050)^\ddag$ \\
  \hline
\end{tabular}
\end{center}
\vspace{-.45cm} \caption{Spectroscopic notation
$n^{2S+1}L_J$, where $\ddag$ is undetected
experimentally.}
\label{notacao-espectroscopica}
\end{table}
 To include all subprocesses in the results, it is necessary to
multiply   $\Gamma$ by a multiplicity factor
$\cal {F}$. For example, in $\phi\rightarrow KK$,   
the possible subprocesses are: $\phi\rightarrow K^+K^-$ and $\phi\rightarrow
K^0\bar{K}^0$. Therefore,  the multiplicity factor is $\cal{F}$=2.
The values of $\cal {F}$ are  listed in table \ref{F}.
\begin{table}[h]
\begin{center}
\begin{tabular}{ c|c|c }
  Generic decay & example & $\cal{F}$ \\
\hline
  $\phi\rightarrow(n\bar{s})(s\bar{n})$        & $\phi_3(1850)\rightarrow K^+K^-$ & 2 \\
  $\phi\rightarrow(n\bar{s})(s\bar{n})^\prime$ & $\phi(1680)\rightarrow K^+K^{\ast-}$ & 4 \\
  $\phi\rightarrow(m)_{I=0}\,(m)_{I=0}$$^{(1)}$          & $\phi(2050)\rightarrow \eta\phi$ & 1 \\
  \hline
\end{tabular}
\end{center}
\vspace{-.4cm} \caption{Multiplicity factor  $\cal{F}$.}
\label{F}
\end{table}

The masses    were obtained from \cite{pdg}, with
exception of   $\phi(2050)$, $\phi_1(1850)$,
$\phi_2(1850)$ that were extracted from \cite{barnes_strange}:
$M_{\phi(1020)}=1.01945$ {\rm GeV}, $M_{\phi(1680)}=1.680$ {\rm
GeV}, $M_{\phi(2050)}=2.050$ {\rm GeV}, $M_{\phi_1(1850)}=1.850$
{\rm GeV}, $M_{\phi_2(1850)}=1.850$ {\rm GeV},
$M_{\phi_3(1850)}=1.854$ {\rm GeV}, $M_{\eta}=0.54785$ {\rm GeV},
$M_{\eta^\prime}=0.95778$ {\rm GeV}, $M_{K}=0.49367$ {\rm GeV},
$M_{K^*}=0.89166$ {\rm GeV}, $M_{K_1(1270)}=1.272$ {\rm GeV},
$M_{K_1(1400)}=1.403$ {\rm GeV}, $M_{K_0^\ast(1430)}=1.425$ {\rm
GeV}, $M_{K_2^\ast(1430)}=1.4256$ {\rm GeV},
$M_{K^\ast(1410)}=1.414$ {\rm GeV}, $M_{K(1460)}=1.460$ {\rm GeV},
$M_{h_1(1380)}=1.386$ {\rm GeV}.

For the initial or final state mesons the Gaussian width 
parameter is set to it's characteristic value 
used for light mesons, namely $\beta_i=0.4\;\rm{GeV}$
\cite{barnes_strange,prd08}.
The   pair production strength parameter $\gamma$
and the angle $\theta$ in (\ref{k1}) and (\ref{k1bar}) were also set
according to \cite{barnes_strange,prd08} $\gamma=0.4$ and $\theta\simeq
35.3^o$ ( $\cos\theta=\sqrt{2/3},\, \sin\theta=\sqrt{1/3})$. 
The $h_1(1380)$ meson is considered a pure $n\bar{n}$ therefore
the coefficients in (\ref{c1c2}) assume the values $c_1^{h_1}=1/\sqrt{2}$ and
$c_2^{h_1}=0$.
The Gaussian widths $\beta_i$,  will 
 remain as free parameters as well as $c_1^i$ and $c_2^i$ coefficients
in equation (\ref{I0}) for the bound-state kernel.
The  parameters $c_i^{\eta\,(\eta^\prime)}$ and
$c_i^{\eta_\Delta\,(\eta^\prime_\Delta)}$ shall be a functions of the same
mixing angle $\theta_p$.
Similarly with $c_i^{\phi\,(\omega)}$ and $c_i^{\phi_\Delta\,(\omega_\Delta)}$,
which are defined by an angle $\theta_v$. Thus the free parameters
to be adjusted are: $\theta_p$, $\theta_{v}$,
$\theta_{v(1680)}$, $\theta_{v_{\,3}}$, $\beta_{\eta_\Delta}$,
$\beta_{\eta^\prime_\Delta}$, $\beta_{\phi_\Delta}$ and
$\beta_{\omega_\Delta}$. Where  $\theta_{v}$, $\theta_{v(1680)}$ and
$\theta_{v_{\,3}}$ are the mixing angles of
$\phi(1020)$, $\phi(1680)$ and $\phi_3(1850)$, respectively.


\subsubsection{$S$  states}

The  $\phi(1020)$ is a natural candidate for the $s\bar{s}$ state
with  $\phi(1680)$ as radial  excitation. One must note that
the decay of $\phi(1680)$ in $KK$ and $KK^\ast$ 
is sometimes cited as evidence that this state is $s\bar{s}$.
The free parameters, that shall be varied, will be the wave functions width $\beta$ 
and the mixing angles.
The decay of  $\phi(1020)\to K^+K^-$ has a partial width of $2.08\pm0.04$ MeV.
Following  the predicted values for the mixing angles
\cite{pdg}, we varied $\theta_p$ between $-20^o$ and $-10^o$ and
$\theta_v$ between $26^o$ and $35^o$. The $\beta_{i_\Delta}$'s were
varied in the range from $0.3$ to $0.6$  GeV. 

The two best fits for this channel have  values of  $\theta_p=-10^o$,
$\theta_v=35^o$, $\beta_{\eta_\Delta} = 2\beta_{\eta^\prime_\Delta}=      0.6$  GeV ,
which we shall call parameterization $(a)$, resulting in
$\Gamma_{\rm 3P0}=3.21$ MeV and $\Gamma_{\rm C3P0}=2.81$ MeV
A different parameterization, which we shall call $(b)$, has
 $\theta_p=-10^o$, $\theta_v=26^o$, 
$4\beta_{\eta_\Delta}= 3 \beta_{\eta^\prime_\Delta} =1.2 $ GeV, 
resulting in
$\Gamma_{\rm 3P0}=2.38$ MeV and $\Gamma_{\rm C3P0}=2.01$ MeV.


The $\phi(1680)$ meson has a total $\Gamma^{\rm tot}_{\rm exp}=150 \pm 50$ MeV,
the C\3p0 model's best fit yields a  $\Gamma^{\rm tot}_{\rm C3P0}=233.29$ MeV,
which corresponds to the values: $\theta_{v(1680)}=35^o$,
$\beta_{\eta_\Delta}=0.6\;\rm{GeV}$,
$\beta_{\eta^\prime_\Delta}=0.3\;\rm{GeV}$,
$\beta_{\phi_\Delta}=0.4\;\rm{GeV}$ and
$\beta_{\omega_\Delta}=0.6\;\rm{GeV}$. The estimated  channels  are in
table \ref{tabphi1680}.

From these results one can see  that for 
$\phi(1020)$, the decay width is within the  experimental range. The same
does not happen for $\phi(1680)$,  where the total decay width is
above the experimental value. It should be noted that in the literature there
are results that indicate higher experimental values: $211\pm14\pm19$ MeV in \cite{shen} 
and $322\pm77\pm 160$ MeV in \cite{phi1680b}. Another important experimental result are the ratios 
$\Gamma_{KK}/\Gamma_{KK^*}=0.07\pm 0.01$ and $\Gamma_{\eta\phi}/\Gamma_{KK^*}\approx 0.37$ \cite{pdg}. 
C\3p0 model's  fit, yields $\Gamma_{KK}/\Gamma_{KK^*}=0.71$ and $\Gamma_{\eta\phi}/\Gamma_{KK^*}=0.19$.
The $\theta_{v(1680)}$ angle is a  measure  of strangeness content of $\phi(1680)$.
A tentative solution to improve these ratios is to set this angle for values beyond 
the $26^o-35^o$ range. 
An increase the angle can lower the total decay width
into the range of experimental values ​​($150\pm50\,\rm{GeV}$) and
 $\Gamma_{KK}/\Gamma_ {KK^*}$ is also
improved. However the ratio  $\Gamma_{\eta\phi}/\Gamma_{KK^*}$ 
didn't improve. This inconsistency could be an
indication that the composition of $\phi(1680)$  
is not well described  and it may be a mixture of the states
$1^3D_1$ and $2^3S_1$, or may have have hybrid components
\cite{pdg}.

The $\phi(2050)$ is a $3 ^{3}S_1$ $s\bar{s}$ vector state, to which   an estimated
mass of 2.05 GeV is assigned, although this state is actually  not known at present. 
It should be important in future spectroscopic studies because with
$1^{--}$ quantum numbers it can be made both in diffractive
photoproduction and in $e^+e^ -$ annihilation. As a radial excitation
of  $s\bar{s}$ one can  use the previous   parameterizations. 
Due to the fact that $\theta_v$ varies between $26^o$ and $35^o$, we
shall consider the extreme values and consider four sets of
parameters. The results of these 
calculations are presented in table \ref{tabphi2050} and yields an 
average total C\3p0 width of $\Gamma= 182.35 \pm 3.13$ MeV.

\begin{table}[!ht]
\begin{ruledtabular}
\begin{tabular}{ccccc}
  $\Gamma$ &3P0  & C3P0  &  C3P0 & Exp   
\\
     (MeV) & (a) & (a)   & (b)   &
\\
   \hline
   $KK$      & 89.42  & 87.42    & 87.42 &    \\
   $KK^\ast$ &123.28 &  123.25& 122.38 &    \\
   $\eta\phi$&21.63   & 26.81& 23.49 &    \\
   \hline
   $\Gamma^{\rm tot}$ & 234.33& 237.48 & 233.29 & $150\pm 50$ \\
\end{tabular}
\end{ruledtabular}
\caption{Decay width of  $\phi(1680)$ 
with  
$\theta_p=-10^o$,
$\theta_v=26^o$, 
$\theta_{v(1680)}=35^o$,
(a)
$2\beta_{\eta_\Delta}=
4\beta_{\eta^\prime_\Delta}=
3\beta_{\phi_\Delta}=
2\beta_{\omega_\Delta}=1.2$ GeV
and  (b)
$2\beta_{\eta_\Delta}=
4\beta_{\eta^\prime_\Delta}=
4\beta_{\phi_\Delta}=
3\beta_{\omega_\Delta}=1.2$ GeV.
}
\label{tabphi1680}
\end{table}

\begin{table}[!t]
\begin{ruledtabular}
\begin{tabular}{ccccccc}
  $\Gamma$ &3P0   & C3P0  & C3P0  &  C3P0 & C3P0  
\\
  (MeV) &(a) & $\rm{(1)}$ & $\rm{(2)}$ & $ \rm{(3)}$  & $\rm{(4)}$ \\
   \hline
  $K K$            &0.06    & $0.08$  & $0.08$  &$0.06$ &$0.06$   \\
  $K K^\ast$       &9.88   & $7.59$  & $7.60$  &$9.89$ &$9.90$   \\
  $K^\ast K^\ast$  &58.93   & $71.36$ & $67.52$ &$59.27$&$56.08$   \\
  $KK_1(1270)$     &6.55   & $5.90$  & $6.02$  &$6.37$ &$6.51$   \\
  $KK_1(1400)$     &15.03   & $16.51$ & $16.82$ &$14.79$&$15.09$   \\
  $KK_0^\ast(1430)$&0.00   & $0.00$  & $0.00$  &$0.00$ &$0.00$   \\
  $KK_2^\ast(1430)$&2.09   & $2.54$  & $2.47$  &$2.11$ &$2.05$   \\
  $KK^\ast(1410)$  &46.72   & $36.33$ & $34.91$ &$47.34$&$45.50$   \\
  $KK(1460)$       &29.46   & $32.23$ & $33.91$ &$26.77$&$28.17$   \\
  $\eta\phi$       &10.44   & $9.49$  & $9.57$  &$10.11$&$10.20$   \\
  $\eta^\prime\phi$&4.61   & $4.52$  & $4.56$  &$4.48$ &$4.52$   \\
  $\eta h_1(1380)$ & 0.00  & $0.08$  & $0.07$  &$0.00$ &$0.00$   \\
  \hline
  $\Gamma^{\rm tot}$&183.77  & $186.63$& $183.53$&$181.19$&$178.08$   \\
\end{tabular}
\end{ruledtabular}
%
\caption{Decay width of $\phi(2050)$, with   $\theta_p=-10^o$, $\theta_v=26^o$
and  the parameterizations:
\\
(a)
$\theta_{v(2050)}=35^o$, 
$2\beta_{\eta_\Delta}=
4\beta_{\eta^\prime_\Delta}=
4\beta_{\phi_\Delta}=
3\beta_{\omega_\Delta}=1.2$ GeV,
\\(1)
$\theta_{v(2050)}=26^o$, 
$2\beta_{\eta_\Delta}=
4\beta_{\eta^\prime_\Delta}=
3\beta_{\phi_\Delta}=
2\beta_{\omega_\Delta}=1.2$ GeV,
\\(2)
$\theta_{v(2050)}=26^o$,
$2\beta_{\eta_\Delta}=
3\beta_{\eta^\prime_\Delta}=
2\beta_{\phi_\Delta}=
4\beta_{\omega_\Delta}=1.2$ GeV,
\\(3)
$\theta_{v(2050)}=35^o$,
$2\beta_{\eta_\Delta}=
4\beta_{\eta^\prime_\Delta}=
3\beta_{\phi_\Delta}=
2\beta_{\omega_\Delta}=1.2$ GeV,
\\(4)
$\theta_{v(2050)}=35^o$,
$2\beta_{\eta_\Delta}=
3\beta_{\eta^\prime_\Delta}=
2\beta_{\phi_\Delta}=
4\beta_{\omega_\Delta}=1.2$ GeV.
}
 \label{tabphi2050}
 \end{table}

\subsubsection{$D$ states}

As well known $\phi_3(1850)$ was first reported in $K^{-}p\to\phi_3\Lambda$ 
at CERN's bubble-chamber experiment \cite{phi3}.
Originally  reported in $KK$ and $KK^\ast$, with a total width of $87^{+28}_{-23} $ MeV and a
relative branching fraction of $B( KK^\ast/KK) =0.55_{-0.45}^{+0.85}$ \cite{pdg}.
Following the same strategy as in the case of the $S$-states,
the $\phi_3(1850)$ parameters that resulted in the best fit are
shown in table \ref{tabphi3} 
were:
$\theta_p=-10^o$, $\theta_v=26^o$, $\theta_{v_{\,3}}=35^o$,
$\beta_{\eta_\Delta}=0.6\;\rm{GeV}$,
$\beta_{\eta^\prime_\Delta}=0.4\;\rm{GeV}$,
$\beta_{\phi_\Delta}=0.6\;\rm{GeV}$ and
$\beta_{\omega_\Delta}=0.3\;\rm{GeV}$.  
The C\3p0 branching fraction resulted in $B( KK^\ast/KK) =0.13$,
in accordance with the experimental limit.

After fixing these parameters, it was possible to estimate the decay widths for the  unobserved mesons
$\phi_1(1850)$, $\phi_2(1850)$. The estimates are shown in table 
\ref{tabphi2}.
Again four sets of parameters were considered in these calculations.
The  average total C\3p0 width's estimates are $\Gamma(\phi_1)= 0.595 \pm 0.058$ MeV and
$\Gamma (\phi_2)= 182.10 \pm 15.87$ MeV.

\begin{table}[t]
\begin{ruledtabular}
\begin{tabular}{ccccc}
  $\Gamma$ &3P0  & C3P0 & C3P0 & Exp
\\
     (MeV) & (a) & (a)   & (b) &  
\\
%
   \hline
   $KK$           & 46.28  & 46.04  & $45.76$ &    \\
   $KK^\ast$      & 6.02  & 6.02 &$6.08$ &    \\
   $K^\ast K^\ast$& 35.24  & 35.26&$33.58$ &    \\
   $KK_1(1270)$   & 0.98  &0.97 &$0.92$ &    \\
   $\eta\phi$     & 0.68  & 0.84&$0.72$ &    \\
   \hline
   $\Gamma^{\rm tot}$&89.20 &89.03& $87.06$ & $87^{+28}_{-23} $ \\
\end{tabular}
\end{ruledtabular}
\caption{Decay width of $\phi_3(1850)$ 
$\theta_p=-10^o$, $\theta_v=26^o$, $\theta_{v_{\,3}}=35^o$,
 with   
and  the parameterizations:
\\
(a)
$2\beta_{\eta_\Delta}=
4\beta_{\eta^\prime_\Delta}=
4\beta_{\phi_\Delta}=
3\beta_{\omega_\Delta}=1.2$ GeV,
\\
(b)
$2\beta_{\eta_\Delta}=
3\beta_{\eta^\prime_\Delta}=
2\beta_{\phi_\Delta}=
4\beta_{\omega_\Delta}=1.2$ GeV.
}
\label{tabphi3}
\end{table}

\begin{table}[t]
\begin{ruledtabular}
\begin{tabular}{cccccc}
          &3P0   & C3P0 & C3P0 & C3P0 & C3P0
\\
    & (a)& $\rm{(1)}$ & $\rm{(2)}$ & $ \rm{(3)}$  & $\rm{(4)}$ \\
   \hline\\
$\Gamma(\phi_1)$ (MeV)\\
  $K^\ast K^\ast$&0.54   &$0.67$ & $0.63$  &$0.56$&$0.52$ \\
\hline\hline \\
$\Gamma(\phi_2)$ (MeV)\\
   $K K$         & 0.0 & $0.00$   & $0.00$   &$0.00$  &$0.00$   \\
   $K K^\ast$     &135.71 & $104.31$ & $103.83$ &$135.94$&$135.31$   \\
   $K^\ast K^\ast$& 13.81& $16.82$  & $16.32$  &$13.97$ &$13.56$   \\
   $\eta\phi$     &50.74 &$45.14$  & $46.05$  &$48.12$ &$49.05$   \\
  \hline
   $\Gamma^{\rm tot}(\phi_2)$&200.26 & $166.27$ & $166.20$  &$198.03$&$197.92$   \\
\end{tabular}
\end{ruledtabular}
\caption{Decay widths of $\phi_1(1850)$ and $\phi_2(1850)$ with
with   $\theta_p=-10^o$, $\theta_v=26^o$
and  the parameterizations:
\\
(a)
$\theta_{v_2}=35^o$, 
$2\beta_{\eta_\Delta}=
4\beta_{\eta^\prime_\Delta}=
4\beta_{\phi_\Delta}=
3\beta_{\omega_\Delta}=1.2$ GeV,
\\(1)
$\theta_{v_2}=26^o$, 
$2\beta_{\eta_\Delta}=
4\beta_{\eta^\prime_\Delta}=
3\beta_{\phi_\Delta}=
2\beta_{\omega_\Delta}=1.2$ GeV,
\\(2)
$\theta_{v_2}=26^o$,
$2\beta_{\eta_\Delta}=
3\beta_{\eta^\prime_\Delta}=
2\beta_{\phi_\Delta}=
4\beta_{\omega_\Delta}=1.2$ GeV,
\\(3)
$\theta_{v_2}=35^o$,
$2\beta_{\eta_\Delta}=
4\beta_{\eta^\prime_\Delta}=
3\beta_{\phi_\Delta}=
2\beta_{\omega_\Delta}=1.2$ GeV,
\\(4)
$\theta_{v_2}=35^o$,
$2\beta_{\eta_\Delta}=
3\beta_{\eta^\prime_\Delta}=
2\beta_{\phi_\Delta}=
4\beta_{\omega_\Delta}=1.2$ GeV.
 } 
\label{tabphi2}
\end{table}

\section{Comparing $^{3}P_{0}$ and $C^{3}P_{0}$  models}

In our former study, in Ref. \cite{dimi10},
 decays in the light $1S$ and $1P$  sectors were analyzed
and  a comparison was made with the usual \3p0 results.
Specifically, the decay processes:
$\rho\to\pi\pi$, $b_1\to\omega\pi$, $a_1\to\rho\pi$,
$a_2\to\rho\pi$, $h_1\to\rho\pi$, $f_0\to\pi\pi$ and
$f_2\to\pi\pi$. One of the  highlights of this study was the fact that
all of these  channels had experimental data.
The model was adjusted in order  to minimize $R$, defined by
\bea 
R^2=\sum_{i=1}^{7} \left[a_{i}(\gamma,\beta)-1\right]^2 
\label{hiper} 
\eea 
with  $a_{i}(\gamma,\beta)= \Gamma_{i}^{\rm thy}(\gamma,\beta) /\Gamma_{i}^{\rm exp} $.
The comparison  of the \3p0 model with C\3p0 implied in obtaining a minimum value for (\ref{hiper}) 
as a function of the parameters $\gamma$ and $\beta$. It was shown that 
the inclusion of the   correction terms reduced the $R$ value.  A clear demonstration that 
the bound-state correction globally improves the fit.
The average difference between the predictions of \3p0 and C\3p0, in each individual channel, ranged from 1\% to 14\%.
The higher differences were in the channels with lighter mesons in the final state. In the heavier channels, the leading
order \3p0 is dominant and the bound-state corrections represent an actual {\it next to leading order} correction. 

In the present, we studied the strange $S$ and $D$ states where data from individual channels are, in general, still not
known. In the best situation, only the total $\Gamma$ has an experimental value. Again a comparison was made between the 
theoretical predictions for \3p0 and C\3p0. For example, in the decay of $\phi(1020)\to KK$, 
two fits were presented with a  difference of 14\% between  \3p0 and C\3p0. Again, the impact of the correction was larger 
in a channel with lighter mesons.  A qualitative interpretation consists in observing that the tightly bound
quark-antiquark pair of lower states are affected in a larger extent by the bound-state correction,
when compared to their radial  excitation.
In the heavier channels, as seen in
tables \ref{tabphi1680} to \ref{tabphi2}, this effect is clear and
the discrepancy falls again to about
1\%. Similar to the case of the heavier mesons of $1S$ and $1P$ sectors studied in \cite{dimi10}.


\section{Conclusions}
 \label{conc}

In this paper, we have concentrated on the $\phi$ mesons,
which are the strange $S$ and $D$ states,  predicted in the quark model, 
as probable $s\bar{s}$ resonances  expected up to 2.2 GeV.
The central body of this study was to employ the Fock-Tani formalism,
a field-theoretic mapping technique in order
to obtain an effective interaction for meson decay. 
  We have outlined the 
essential aspects of the $C^{3}P_{0}$ model, the bound-state corrected 
\3p0 model, and how it is 
obtained from the Fock-Tani formalism, where the decay
Hamiltonian $H^{C3P0}$ was deducted. 

This work is intended as a modest guide for future experimental
studies of meson spectroscopy that may  become possible with
the  advent of the new Hall D photoproduction facility GlueX at
Jefferson Lab.  The main goal of the GlueX experiment is to search 
for and study hybrid and exotic mesons. In this sense we have studied 
6  $s\bar{s}$  states, in which 3 were unobserved, presenting interesting 
issues for future experimental studies involving the conventional quark
model states.  The predicted total
widths for these new, rather narrow states, that have not been identified,  are
$\Gamma(\phi(2050))= 182.35 \pm 3.13$ MeV, $\Gamma(\phi_1)= 0.595 \pm 0.058$ MeV and
$\Gamma (\phi_2)= 182.10 \pm 15.87$ MeV.


\acknowledgements

This research was supported by
Conselho Nacional de Desenvolvimento Cient\'{\i}fico e Tecnol\'ogico (CNPq),
Universidade Federal do Rio Grande do Sul (UFRGS).
and Universidade Federal de Pelotas (UFPel).


\appendix

\section{Physical nature of  bound-state corrections}

The bound-state corrections   are an essential aspect of the Fock-Tani formalism, sometimes
called {\it orthogonality corrections} because they are equivalent to the description of the continuum states 
by orthogonalized plane waves   with the projection to bound-states subtracted \cite{girar1}-\cite{girar3}.
Consider, for example, an ideal   two meson state (which shall be represented by a round ket)
\bea
|\alpha\beta)=m^\dag_\alpha\,m^\dag_\beta\,\ket 0\,.
\label{phy1}
\eea 
The norm of (\ref{phy1}) can be calculated using (\ref{mcom}):
\bea
(\rho\sigma|\alpha\beta)=
\delta_{\alpha\,\rho}\,\delta_{\beta\,\sigma} +
\delta_{\alpha\,\sigma}\,\delta_{\beta\,\rho} \,.
\label{phy2}
\eea
The same calculation can be done for the physical  two meson state
\bea
\ket{\alpha\beta}=M^\dag_\alpha\,M^\dag_\beta\,\ket 0\,.
\label{phy3}
\eea
Defining the norm function as $N(\rho\sigma;\alpha\beta)\equiv\bk{\rho\sigma}{\alpha\beta}$
and using (\ref{M-com}), one
obtains
\bea
N(\rho\sigma;\alpha\beta)=
\delta_{\alpha\,\rho}\,\delta_{\beta\,\sigma} - N_E(\rho\sigma;\alpha\beta)
+(\rho\leftrightarrow\sigma)
\nn\\
\label{phy4}
\eea
where
\bea
N_E(\rho\sigma;\alpha\beta)=
\Phi^{\ast\,\xi\tau}_{\rho}
\Phi^{\ast\,\omega\lambda}_{\sigma}
\Phi^{\xi\lambda}_{\alpha}
\Phi^{\omega\tau}_{\beta}\,.
\label{phy5}
\eea
The presence of $N_E$ in (\ref{phy4}) has its origin in the composite nature of the meson operator $M_\alpha$
and implies that the two meson state is not normalized as in (\ref{phy2}).
A correctly normalized state would be written as
\bea
\overline{\ket{\alpha\beta}}=N^{-\frac{1}{2}}(\alpha\beta;\alpha^\prime\beta^\prime) \,\ket{\alpha^\prime\beta^\prime}\,.
\label{phy6}
\eea
Now, consider the following state
\bea
\ket{\mu\nu\alpha}=q^\dag_\mu\,\bar{q}^\dag_\nu\,M^\dag_\alpha\,\ket 0\,,
\label{phy7}
\eea
which by a similar procedure  can be normalized, resulting in
\bea
\overline{\ket{\mu\nu\alpha}}=N^{-\frac{1}{2}}_q(\mu\nu\alpha;\mu^\prime\nu^\prime\alpha^\prime) 
\ket{\mu^\prime\nu^\prime\alpha^\prime}\,,
\label{phy8}
\eea
where
\bea
\hspace{-1cm}
N_q(\mu\nu\alpha;\mu^\prime\nu^\prime\alpha^\prime)&=&
\delta_{\mu\mu^\prime}\delta_{\nu\nu^\prime}\delta_{\alpha\alpha^\prime}
-N_{E}^q(\mu\nu\alpha;\mu^\prime\nu^\prime\alpha^\prime)
\label{phy9}
\eea
with
\bea
N_{E}^q(\mu\nu\alpha;\mu^\prime\nu^\prime\alpha^\prime)&=&
\delta_{\mu\mu^\prime}\,\Phi^{\ast\xi\nu}_{\alpha^\prime}\,\Phi^{\xi\nu^\prime}_{\alpha}
+\delta_{\nu\nu^\prime}\,\Phi^{\ast\mu\tau}_{\alpha^\prime}\,\Phi^{\mu^\prime\tau}_{\alpha}
\nn\\&&
-\,\Phi^{\ast\mu\nu}_{\alpha^\prime}\,\Phi^{\mu^\prime\nu^\prime}_{\alpha}\,.
\label{phy9b}
\eea

A decay in which $A\to B C$  is described by an amplitude obtained evaluating the following matrix element  
$\bra{BC}V\ket A$. In second quantization this is written as
\bea
\bra{BC}V\ket A &=& \bra{0} M_{\alpha}M_{\beta} V_{\mu\nu}q^\dag_\mu\,\bar{q}^\dag_\nu\,M^\dag_\gamma\,\ket 0\,
\nn\\
&=&
V_{\mu\nu}\,\bk{\alpha\beta}{\mu\nu\gamma}\,.
\label{phy10}
\eea
According to what was shown, the state vectors in the decay amplitude (\ref{phy10}) should be replaced by a normalized version
\bea
\overline{\bra{BC}V {\ket A}}&=&
V_{\mu\nu}\,\overline{\bk{\alpha\beta}{\mu\nu\gamma}}
\nn\\
&=&
V_{\mu\nu}\,\
N^{-\frac{1}{2}}(\alpha\beta;\beta^\prime\alpha^\prime) 
\bk{\alpha^\prime\beta^\prime}{\mu^\prime\nu^\prime\gamma^\prime}
\nn\\&&\times
N^{-\frac{1}{2}}_q(\mu^\prime\nu^\prime\gamma^\prime;\mu\nu\gamma)\,.
\label{phy11}
\eea
Defining a potential norm function as 
\bea
N_V(\mu\nu\alpha^\prime\beta^\prime;\mu^\prime\nu^\prime\gamma^\prime ) 
&\equiv&  V_{\mu\nu} \bk{\alpha^\prime\beta^\prime}{\mu^\prime\nu^\prime\gamma^\prime}
\nn\\
&=&V_1-V_3
\label{phy12}
\eea
where $V_{i}$, with one or three wave functions, is given by
\bea
V_1 &=&\Phi^{\ast\mu^\prime\nu^\prime}_{\alpha^\prime}\delta_{\beta^\prime\gamma^\prime}V_{\mu\nu} +
(\alpha^\prime\leftrightarrow\beta^\prime)
\nn\\
V_3 &=&
\Phi^{\ast\mu^\prime\tau}_{\alpha^\prime}
\Phi^{\ast\omega\nu^\prime}_{\beta^\prime}
\Phi^{\omega\tau}_{\gamma^\prime}
V_{\mu\nu}
+(\alpha^\prime\leftrightarrow\beta^\prime)\,.
\label{phy13}
\eea
The complete evaluation of the  normalized decay amplitude is reduced to the expansion of
(\ref{phy11}):
\bea
\overline{\bra{BC}V {\ket A}}
&=&
N^{-\frac{1}{2}}(\alpha\beta;\beta^\prime\alpha^\prime) 
N_V(\mu\nu\alpha^\prime\beta^\prime;\mu^\prime\nu^\prime\gamma^\prime ) 
\nn\\&&\times
N^{-\frac{1}{2}}_q(\mu^\prime\nu^\prime\gamma^\prime;\mu\nu\gamma)\,.
\label{phy14}
\eea
For example, in the lowest order of the expansion  
\bea
N^{-\frac{1}{2}}&\approx& 1+\frac{1}{2}N_E
\hs
N_q^{-\frac{1}{2}}\approx 1+\frac{1}{2}N_E^q
\label{phy14b}
\eea
one obtains
\bea
\overline{\bra{BC}V {\ket A}}
&\approx&
-\Phi^{\ast\rho\xi}_{\alpha} 
\Phi^{\ast\lambda\tau}_{\beta}
\Phi^{\omega\sigma}_{\gamma}\,
V(\rho\xi\lambda\tau;\omega\sigma)
\nn\\&&
+(\alpha \leftrightarrow\beta)\,,
\label{phy15}
\eea
where
\bea
V(\rho\xi\lambda\tau;\omega\sigma)&=&\left(\,\bar{\delta} +\bar{\Delta}_{f}\,\right)\,V_{\mu\nu}\,,
\label{phy16}
\eea
with
\bea
\bar{\delta}&=&\delta_{\mu\lambda}
\delta_{\nu\xi}
\delta_{\omega\rho}
\delta_{\sigma\tau}
\nn\\
\bar{\Delta}_f&=&
f\,\left[
\delta_{\sigma\xi}\,\delta_{\lambda\mu}\,\,
\Delta(\rho\tau;\omega\nu)
\frac{}{}
+
\delta_{\xi\nu}\,\delta_{\lambda\omega}\,\,
\Delta(\rho\tau;\mu\sigma)
\right.
\nn\\
&&
\left. \frac{}{}
-2\delta_{\sigma\xi}\,\delta_{\lambda\omega}\,\,
\Delta(\rho\tau;\mu\nu)
\right]\,.
\label{phy17}
\eea
In (\ref{phy17}), $f$ is combinatorial factor with the value $f=1$ related to the truncation of 
(\ref{phy14b}) in the lowest order. When higher orders are considered new contributions   change
this factor to the Fock-Tani value $f=1/4$ of (\ref{deltas}).

In summary, the essence of the Fock-Tani formalism is to move the bound-state information from
the state vector (\ref{1b}), written in second quantization, into the interaction  (\ref{c3p0}).
As shown in the {\it First} and {\it Second examples}, this action has different impacts in the
physical system.  The new state vector is now 
the ideal state vector (\ref{single_mes}) which satisfies canonical commutation relations (\ref{mcom}).

If one chooses not to use the Fock-Tani formalism the decay amplitude can be evaluated directly  by calculating the 
$\bra{BC}V\ket A$ matrix element. As a first approximation, neglecting the meson's composite structure, the \3p0
gives a correct  leading order contribution to $\bra{BC}V\ket A$. To go beyond this lowest order, one needs to calculate 
$\overline{\bra{BC}V {\ket A}}$ and expand the kernels $N$, $N_q$ and $N_V$ in (\ref{phy14}), introducing the overlap 
effects due to the extended nature of the meson, which   constitute the bound-state corrections.

\section{Wave functions}
\label{funcao-onda}

The Pauli matrices are given by
\begin{equation}
\sigma_{x}=\left(\begin{array}{cr}
0 & 1\\
1 & 0\end{array}\right);\;\sigma_{y}=\left(\begin{array}{cr}
0 & -i\\
i & 0\end{array}\right);\;\sigma_{z}=\left(\begin{array}{cr}
1 & 0\\
0 & -1\end{array}\right)
\label{e1}
\end{equation}
and the corresponding spinors are
\begin{equation}
\chi_{1}=\left(\begin{array}{c}
1\\
0\end{array}\right);\;\chi_{2}=\left(\begin{array}{c}
0\\
1\end{array}\right);\;\chi_{1}^{c}=\left(\begin{array}{c}
0\\
1\end{array}\right);\;\chi_{2}^{c}=\left(\begin{array}{c}
-1\\
0\end{array}\right)\;.
\label{e2}
\end{equation}
The  meson flavor components listed in the decay channels (a)-(f) depend on the isospin $I$ and strangeness
$s$

\begin{enumerate}
\item $ \underline{I=0}$
\begin{equation}
\phi,\, \eta,\, \eta^\prime, \,h_1\;\rightarrow\,\frac{c_{1}}{\sqrt{2}}\left(|u\bar{u}
 \ran +|d\bar{d} \ran \right)+c_2\,|s\bar{s} \ran
\label{I0}
\end{equation}

\item $\underline{I=1/2,\, s=+1}$
 \begin{equation}
\begin{array}{c}
K^+\;\rightarrow\,-|u\bar{s}\ran \quad ; \quad I_z=+1/2\\
K^0\;\rightarrow\,-|d\bar{s}\ran \quad ; \quad I_z=-1/2
\end{array}
\label{I12a}
\end{equation}
 
\item $\underline{I=1/2,\, s= -1}$
\begin{equation}
\begin{array}{c}
\bar{K}^0\;\rightarrow\,-|s\bar{d}\ran \quad ; \quad I_z=+1/2\\
K^-\;\rightarrow\,|s\bar{u}\ran \quad ; \quad I_z=-1/2\;.
\end{array}
\label{I12b}
\end{equation}
\end{enumerate}

The $SU(3)$ mixing coefficients $c_{1}$ and $c_{2}$, in (\ref{I0}), for the pseudo-scalar mesons $\eta$ and
$\eta^\prime$ are related through the angle $\theta_p$
\bea
|\eta\rangle &=&\frac{c_{1}(\theta_{p})}{\sqrt{2}}
\left[ |u\bar{u}\rangle +  |d\bar{d}\rangle \,\right]- c_{2}(\theta_{p})|s\bar{s}\rangle
\nn\\
|\eta^{\prime}\rangle &=& 
\frac{c_{2}(\theta_{p})}{\sqrt{2}}\left[ |u\bar{u}\rangle +  |d\bar{d}\rangle \,\right]
+ c_{1}(\theta_{p}) |s\bar{s}\rangle\,.
\label{eta-etal}
\eea
Similarly, the coefficients of the vector mesons $\phi(1020)$ and $\omega(782)$ are related through
angle $\theta_v$.
\bea
|\phi  \rangle &=& \frac{c_{1}(\theta_{v})}{\sqrt{2}}
\left[ |u\bar{u}\rangle +  |d\bar{d}\rangle \,\right]- c_{2}(\theta_{v})|s\bar{s}\rangle
\nn\\
|\omega\rangle &=&
 \frac{c_{2}(\theta_{v})}{\sqrt{2}}\left[ |u \bar{u}\rangle +  |d \bar{d}\rangle \,\right]
+ c_{1}(\theta_{v}) |s\bar{s}\rangle\,,
\label{phi-w_2} 
\eea 
where
\bea
c_{1}(\theta_{i})&=&\frac{\cos\theta_{i}}{\sqrt{3}}- \sqrt{\frac{2}{3} } \sin\theta_{i} 
\nn\\
c_{2}(\theta_{i})&=&\sqrt{\frac{2}{3}}\cos\theta_{i}+\frac{\sin\theta_i}{\sqrt{3}}\,.
\label{c1c2}
\eea
Equations (\ref{I12a})-(\ref{I12b}) are also valid
mesons to $K_1$, $K^{*}$, $K_0^*$ and $K_2^*$. 
In particular,  $K_1(1270)$ and $K_1(1400)$ are related by mixing angle $\theta$
\bea
\left|K_1(1270)\right\rangle&=& +\cos \theta\, \left|1^1P_1\right\rangle+\sin\theta\,\left|1^3P_1\right\rangle\nn\\
\left|K_1(1400)\right\rangle&=& -\sin\theta
\,\left|1^1P_1\right\rangle+\cos\theta\,\left|1^3P_1\right\rangle\;.
\label{k1} 
\eea 
For antikaons there is a change in the sign 
\bea
\left|\bar{K}_1(1270)\right\rangle&=& -\cos \theta\, \left|1^1P_1\right\rangle+\sin\theta\,\left|1^3P_1\right\rangle\nn\\
\left|\bar{K}_1(1400)\right\rangle&=& +\sin\theta
\,\left|1^1P_1\right\rangle+\cos\theta\,\left|1^3P_1\right\rangle\;.
\label{k1bar} 
\eea
The color wave function is the same for all mesons, is given by
\bea
C^{c_{\mu}c_{\nu}}=\frac{1}{\sqrt{3}}\,\,\delta^{c_{\mu}c_{\nu}}\quad;\quad
c_{k}=1, 2, 3\,. 
\label{color} 
\eea
The spin wave functions can be singlet or triplet:

\begin{itemize}
\item Singlet ($S=0$)
\bea \frac{1}{\sqrt{2}}\left(|\uparrow
\downarrow \hspace{0.2cm} \ran -|\downarrow \uparrow \hspace{0.2cm}
\ran \right)\quad;\quad S_{z}=0
\label{apen-s0} 
\eea

\item Triplet ($S=1$)
\bea
 &|\uparrow \uparrow     \hspace{0.2cm} \ran  &;\quad S_z=+1\nn\\
&\frac{1}{\sqrt{2}}\left(|\uparrow \downarrow \hspace{0.2cm} \ran
+|\downarrow \uparrow \hspace{0.2cm} \ran\right)&; \quad S_{z}=0\nn\\
&| \downarrow \downarrow    \hspace{0.2cm} \ran &;\quad S_{z}=-1\,.
\label{apen-s3} 
\eea
\end{itemize}
The spatial   wave functions are
harmonic oscillator functions, as they describe  color confinement 
and provide analytical amplitudes
 \bea \Phi_{nl
}(\vec{P}_{\alpha}-\vec{p}_{\mu}-\vec{p}_{\nu}) =
\delta(\vec{P}_{\alpha}-\vec{p}_{\mu}-\vec{p}_{\nu})
\,\,\phi_{nl}(\vec{p}_\mu-\vec{p}_\nu)\,, \nn\\
\eea
where
$\phi_{nl}(\vec{p}_\mu-\vec{p}_\nu)  $ is given by
\bea
\phi_{nl}(\vec{p}_\mu-\vec{p}_\nu)&=&
 \left(\frac{1}{2\beta}\right)^{l}\, N_{nl}\, |\vec{p}_\mu-\vec{p}_\nu|^{l}\,
\phi(\vec{p}_\mu-\vec{p}_\nu)
%
\nn\\
&&\times {\cal L}_{n-1}^{l+\frac{1}{2}}
\left[\frac{(\vec{p}_\mu-\vec{p}_\nu)^2   }{ 4\beta^2} \right]
 Y_{lm}(\Omega_{\vec{p}_\mu-\vec{p}_\nu})\,,\nn\\
\label{psi_oh}
\eea
where $p_{\mu(\nu)} $ is the internal momentum, $Y_{lm}$ 
spherical harmonic and $\beta$ the  Gaussian width parameter.
The momentum wave function $\phi(\vec{p}_\mu-\vec{p}_\nu)$, the normalization constant
 $N_{nl}$ and the Laguerre polynomials ${\cal L}_{n-1}^{l+\frac{1}{2}}(p)$, that 
depend on the radial $n$ and orbital $l$ quantum numbers, are all defined as
\bea 
N_{nl}&=& \left[ \frac{2\,
(n-1)!}{\beta^3\,\Gamma(n+l+1/2) } \right]^{\frac{1}{2}} 
\nn\\
{\cal L}_{n-1}^{l+\frac{1}{2}}(p)&=&\sum_{k=0}^{n} \frac{(-)^k\,
\Gamma(n+l+1/2)}{ k!\,(n-k-1)!\, \Gamma(k+l+3/2) }\,\,p^k\,,
\nn\\
\phi(\vec{p}_\mu-\vec{p}_\nu)&=&\exp\left[ -\frac{(\vec{p}_\mu-\vec{p}_\nu)^2   }{ 8\beta^2 }\right]\,
\label{laguerre} 
\eea
where $n=1,2,\ldots$ and $l=0,1,\ldots$


\section{Amplitudes}
\label{amplitudes}

\noindent

In this appendix, we present the results of the algebraic decay
amplitudes of $h_{fi}$, for subprocesses of (a)-(f),
obtained with the C\3p0 model.
Defining
\begin{widetext}
\begin{eqnarray}
e_1(p,\beta_A,\beta_B,\beta_C)&=&e^{-\frac{\left(\beta_B^2+\beta_C^2\right)
p^2}{8 \left(\beta_C^2 \beta_A^2+\beta_B^2
\left(\beta_C^2+\beta_A^2\right)\right)}}\nn\\
e_2(p,\beta_A,\beta_B,\beta_C,\beta)&=& e^{-\frac{\left(\left(2
\beta_A^2+\beta^2\right) \left(\beta_B^2+2 \beta^2\right)+\beta_C^2
\left(2 \left(\beta_A^2+\beta_B^2\right)+5 \beta^2\right)\right)
p^2}{8 \beta^2 \left(\beta_C^2 \left(\beta_B^2+2
\beta^2\right)+\beta_A^2 \left(\beta_B^2+\beta_C^2+2
\beta^2\right)\right)}} \;,
\label{hfic3po-e1}
\end{eqnarray}
\end{widetext}
where $\beta_A$ is the width of the Gaussian initial
state of the meson, $\beta_B$ and $\beta_C$ of the final state
mesons and $\beta$  for the bound-state correction mesons, one has
\begin{eqnarray}
\calm_{fi}^{\phi(1020)\to KK} &=& {\cal C}^{\phi(1020)}_{10}\,
Y_{11}\left(\Omega_p\right)
\label{hfic3po-phi1020a} 
\end{eqnarray}
\begin{eqnarray}
\calm_{fi}^{\phi(1680)\to KK} &=& {\cal C}^{\phi(1680)KK}_{10}\,
Y_{11} \left(\Omega_p\right)
\,\,\,\,\,\,\,\,\,\,\,\,\,\,\,\,\,\,\,\,\,\,\,\,
\label{hfic3po-phi1680kk}
\end{eqnarray}
\begin{eqnarray}
\calm_{fi}^{\phi(1680)\to KK^*} &=& {\cal C}^{\phi(1680)KK^*}_{11}\,
Y_{10} \left(\Omega_p\right)
\,\,\,\,\,\,\,\,\,\,\,\,\,\,\,\,\,\,\,\,\,\,\,\,
\label{hfic3po-phi1680kke}
\end{eqnarray}
\begin{eqnarray}
\calm_{fi}^{\phi(1680)\to \eta\phi} &=& {\cal
C}^{\phi(1680)\eta\phi}_{11}\, Y_{10} \left(\Omega_p\right)
\,\,\,\,\,\,\,\,\,\,\,\,\,\,\,\,\,\,\,\,\,\,\,\,
\label{hfic3po-phi1680etaphi}
\end{eqnarray}
\begin{eqnarray}
\calm_{fi}^{\phi(2050)\to KK} &=& {\cal C}^{\phi(2050)KK}_{10}\,
Y_{11} \left(\Omega_p\right)
\,\,\,\,\,\,\,\,\,\,\,\,\,\,\,\,\,\,\,\,\,\,\,\,
\label{hfic3po-phi2050kk}
\end{eqnarray}
\begin{eqnarray}
\calm_{fi}^{\phi(2050)\to KK^\ast} &=& {\cal
C}^{\phi(2050)KK^\ast}_{11}\, Y_{10} \left(\Omega_p\right)
\,\,\,\,\,\,\,\,\,\,\,\,\,\,\,\,\,\,\,\,\,\,\,\,
\label{hfic3po-phi2050kke}
\end{eqnarray}
\begin{eqnarray}
\calm_{fi}^{\phi(2050)\to K^{\ast}K^{\ast}} &=& {\cal
C}^{\phi(2050)K^{\ast}K^{\ast}}_{12}\, Y_{1-1} \left(\Omega_p\right)
\,\,\,\,\,\,\,\,\,\,\,\,\,\,\,\,\,\,\,\,\,\,\,\,
\label{hfic3po-phi2050keke}
\end{eqnarray}
\begin{eqnarray}
\calm_{fi}^{\phi(2050)\to KK_1(1270)} &=& {\cal
C}^{\phi(2050)KK_1(1270)}_{01}\, Y_{00}
\left(\Omega_p\right)\nn\\
&+&{\cal C}^{\phi(2050)KK_1(1270)}_{21}\, Y_{20}
\left(\Omega_p\right) 
\label{hfic3po-phi2050kk11270}
\nn\\
\end{eqnarray}
\begin{eqnarray}
\calm_{fi}^{\phi(2050)\to KK_1(1400)} &=& {\cal
C}^{\phi(2050)KK_1(1400)}_{01}\, Y_{00}
\left(\Omega_p\right)\nn\\
&+&{\cal C}^{\phi(2050)KK_1(1400)}_{21}\, Y_{20}
\left(\Omega_p\right)
\label{hfic3po-phi2050kk11400}
\nn\\
\end{eqnarray}
\begin{eqnarray}
\calm_{fi}^{\phi(2050)\to KK_0^\ast(1430)} &=& {\cal
C}^{\phi(2050)KK_0^\ast(1430)}_{20}\, Y_{21} \left(\Omega_p\right)
\label{hfic3po-phi2050kkoe1430a}
\nn\\
\end{eqnarray}
\begin{eqnarray}
\calm_{fi}^{\phi(2050)\to KK_2^\ast(1430)} &=& {\cal
C}^{\phi(2050)KK_0^\ast(1430)}_{22}\, Y_{2-1} \left(\Omega_p\right)
\label{hfic3po-phi2050kkoe1430b}
\nn\\
\end{eqnarray}
\begin{eqnarray}
\calm_{fi}^{\phi(2050)\to KK^\ast(1410)} &=& {\cal
C}^{\phi(2050)KK^\ast(1410)}_{11}\, Y_{10} \left(\Omega_p\right)
\,\,\,\,\,\,\,\,\,\,\,\,\,\,
\label{hfic3po-phi2050kke1410}
\nn\\
\end{eqnarray}
\begin{eqnarray}
\calm_{fi}^{\phi(2050)\to KK(1460)} &=& {\cal
C}^{\phi(2050)KK(1460)}_{10}\, Y_{11} \left(\Omega_p\right)
\,\,\,\,\,\,\,\,\,\,\,\,\,
\label{hfic3po-phi2050kk1460}
\end{eqnarray}
\begin{eqnarray}
\calm_{fi}^{\phi(2050)\to \eta\phi} &=& {\cal
C}^{\phi(2050)\eta\phi}_{11}\, Y_{10} \left(\Omega_p\right)
\,\,\,\,\,\,\,\,\,\,\,\,\,\,\,\,\,\,\,\,\,\,\,\,
\label{hfic3po-phi2050etaphi}
\end{eqnarray}
\begin{eqnarray}
\calm_{fi}^{\phi(2050)\to \eta^\prime\phi} &=& {\cal
C}^{\phi(2050)\eta^\prime\phi}_{11}\, Y_{10} \left(\Omega_p\right)
\,\,\,\,\,\,\,\,\,\,\,\,\,\,\,\,\,\,\,\,\,\,\,\,
\label{hfic3po-phi2050etalphi}
\end{eqnarray}
\begin{eqnarray}
\calm_{fi}^{\phi(2050)\to \eta h_1(1380)} &=& {\cal
C}^{\phi(2050)\eta h_1}_{01}\, Y_{00}
\left(\Omega_p\right)\nn\\
&+& {\cal C}^{\phi(2050)\eta h_1}_{21}\, Y_{20}
\left(\Omega_p\right)
\label{hfic3po-phi2050etah1}
\end{eqnarray}
\begin{eqnarray}
\calm_{fi}^{\phi_1 \to K^\ast K^\ast} &=& {\cal C}^{\phi_1 K^\ast
K^\ast}_{12}\, Y_{1-1} \left(\Omega_p\right)\nn\\
&+& {\cal C}^{\phi_1 K^\ast K^\ast}_{32}\, Y_{3-1}
\left(\Omega_p\right)
\label{hfic3po-phi1keke}
\end{eqnarray}
\begin{eqnarray}
\calm_{fi}^{\phi_2 \to K K} &=& {\cal C}^{\phi_2 K K}_{30}\, Y_{32}
\left(\Omega_p\right)
\label{hfic3po-phi2kk}
\end{eqnarray}
\begin{eqnarray}
\calm_{fi}^{\phi_2 \to K K^\ast} &=& {\cal C}^{\phi_2 K
K^\ast}_{11}\, Y_{11} \left(\Omega_p\right)\nn\\
&+&{\cal C}^{\phi_2 K K^\ast}_{31}\, Y_{31}
\left(\Omega_p\right) 
\label{hfic3po-phi2kke}
\end{eqnarray}
\begin{eqnarray}
\calm_{fi}^{\phi_2 \to K^\ast K^\ast} &=& {\cal C}^{\phi_2 K^\ast
K^\ast}_{12}\, Y_{10} \left(\Omega_p\right)\nn\\
&+&{\cal C}^{\phi_2 K^\ast K^\ast}_{32}\, Y_{30}
\left(\Omega_p\right) 
\label{hfic3po-phi2keke}
\end{eqnarray}
\begin{eqnarray}
\calm_{fi}^{\phi_2 \to \eta \phi} &=& {\cal C}^{\phi_2 \eta
\phi}_{11}
\, Y_{11} \left(\Omega_p\right)\nn\\
&+&{\cal C}^{\phi_2 \eta \phi}_{31}\, Y_{31}
\left(\Omega_p\right) 
\label{hfic3po-phi2nphi}\\
\nn
\end{eqnarray}
\begin{eqnarray}
\calm_{fi}^{\phi_3 \to K K} &=& {\cal C}^{\phi_3 K K}_{30}\, Y_{33}
\left(\Omega_p\right) 
\label{hfic3po-phi3kk}
\end{eqnarray}
\begin{eqnarray}
\calm_{fi}^{\phi_3 \to K K^\ast} &=& {\cal C}^{\phi_3 K
K^\ast}_{31}\, Y_{32}
\left(\Omega_p\right) 
\label{hfic3po-phi3kke}
\end{eqnarray}
\begin{eqnarray}
\calm_{fi}^{\phi_3 \to K^\ast K^\ast} &=& {\cal C}^{\phi_3 K^\ast
K^\ast}_{12}\, Y_{11} \left(\Omega_p\right)\nn\\
&+&{\cal C}^{\phi_3 K^\ast K^\ast}_{32}\, Y_{31}
\left(\Omega_p\right) 
\label{hfic3po-phi3keke}
\end{eqnarray}
\begin{eqnarray}
\calm_{fi}^{\phi_3 \to K K_1(1270)} &=& {\cal C}^{\phi_3 K
K_1(1270)}_{21}
\, Y_{22} \left(\Omega_p\right)\nn\\
&+&{\cal C}^{\phi_3 K K_1(1270)}_{41}\, Y_{42}
\left(\Omega_p\right) 
\label{hfic3po-phi3kk1}
\end{eqnarray}
\begin{eqnarray}
\calm_{fi}^{\phi_3 \to \eta \phi} &=& {\cal C}^{\phi_3 \eta
\phi}_{31}\, Y_{32} \left(\Omega_p\right) 
\label{hfic3po-phi3nphi}
\end{eqnarray}
where ${\cal C}_{LS}$ are give by

\begin{widetext}
$\rm{\underline{\phi(1020)\rightarrow K^+ K^-}:}$

\begin{eqnarray}
{\cal C}^{\phi(1020)}_{10}&=& (c_1^{\phi}-c_2^{\phi})
\left\{\frac{}{}f_1(p_{KK},\beta_\phi,\beta_K)\,e_1(p_{KK},\beta_\phi,\beta_K,\beta_K)
\right.
\nn\\
&& \left. -  c_1^{\eta_\Delta}
c_2^{\eta_\Delta}\,f_2(p_{KK},\beta_\phi,\beta_K,\beta_{\eta_\Delta})\,
e_2(p_{KK},\beta_\phi,\beta_K,\beta_K,\beta_{\eta_\Delta}) \right.
\nn\\
&&\left. - c_1^{\eta_\Delta^\prime} c_2^{\eta_\Delta^\prime}
f_2(p_{KK},\beta_\phi,\beta_K,\beta_{\eta_\Delta^\prime})\,
e_2(p_{KK},\beta_\phi,\beta_K,\beta_K,\beta_{\eta_\Delta^\prime})\frac{}{}
\right\}
\label{hfic3po-phi1020b}
\end{eqnarray}
\begin{eqnarray} f_1(p,\beta_A,\beta_B)&=& \frac{8\,p\,\beta_A^{3/2}
\left(\beta_B^2+\beta_A^2\right)}{ \sqrt{3}\left(\beta_B^2+2
\beta_A^2\right)^{5/2}}
\label{hfic3po-f1}
\end{eqnarray}
\begin{eqnarray}
f_2(p,\beta_A,\beta_B,\beta)&=&\frac{16\,p\,\beta_A^{3/2}\beta_B^3
\left(\beta_B^2+\beta_A^2\right) \left(\beta_B^2+\beta^2\right) }{3
\sqrt{3}\left(\beta_B^4+2 \beta^2 \beta_A^2+2 \beta_B^2
\left(\beta^2+\beta_A^2\right)\right)^{5/2}} 
\label{hfic3po-f2}
\end{eqnarray}

$\rm{\underline{\phi(1680)\rightarrow K^+ K^-}:}$

\begin{eqnarray}
{\cal C}^{\phi(1680)KK}_{10} &=&
(c_1^{\phi_{1680}}-c_2^{\phi_{1680}})
\left\{\frac{}{}f_3(p_{KK},\beta_{\phi_{1680}},\beta_K)
\,e_1(p_{KK},\beta_{\phi_{1680}},\beta_K,\beta_K) \right.
\nn\\
&&\left. +c_1^{\eta_\Delta}
c_2^{\eta_\Delta}\,f_4(p_{KK},\beta_{\phi_{1680}},\beta_K,\beta_{\eta_\Delta})
\,e_2(p_{KK},\beta_{\phi_{1680}},\beta_K,
\beta_K,\beta_{\eta_\Delta}) \right.
\nn\\
&&\left.+c_1^{\eta_\Delta^\prime}
c_2^{\eta_\Delta^\prime}f_4(p_{KK},\beta_{\phi_{1680}},\beta_K,\beta_{\eta_\Delta^\prime})
\,e_2(p_{KK},\beta_{\phi_{1680}},\beta_K,\beta_K,\beta_{\eta_\Delta^\prime})
\frac{}{}\right\}
\label{hfic3po-phi1680-1}
\end{eqnarray}
\begin{eqnarray}
f_3(p,\beta_A,\beta_B)&=&\frac{4 \sqrt{2}\,p\,\beta_A^{3/2} \left[-3
\beta_B^6+\beta_B^4 \beta_A^2+20 \beta_B^2 \beta_A^4+12 \beta_A^6-2
\beta_A^2 \left(\beta_B^2+\beta_A^2\right)
p^2\right]}{3\left(\beta_B^2+2 \beta_A^2\right)^{9/2}}
\label{hfic3po-f3}
\end{eqnarray}
\begin{eqnarray}
f_4(p,\beta_A,\beta_B,\beta)&=& \frac{8\sqrt{2}\,p \,\beta_B^3\,
\beta_A^{3/2}\left(\beta_B^2+\beta^2\right) }{9\left(\beta_B^4+2
\beta^2 \beta_A^2+2 \beta_B^2
\left(\beta^2+\beta_A^2\right)\right)^{9/2}}\left[\frac{}{}\left(3
\beta_B^4 \left(\beta_B^2+2 \beta^2\right)
 -7 \beta_B^4 \beta_A^2-6 \left(\beta_B^2+\beta^2\right)
\beta_A^4\right) \right.
\nn\\
&&\left. \left(\beta_B^4+2 \beta^2 \beta_A^2 +2
\beta_B^2\left(\beta^2+\beta_A^2\right)\right)+2
\left(\beta_B^2+\beta^2\right)^2 \beta_A^2
\left(\beta_B^2+\beta_A^2\right) p^2\right] 
\label{hfic3po-f4}
\end{eqnarray}

$\rm{\underline{\phi(1680)\rightarrow K^+ K^{\ast-}}:}$

\begin{eqnarray}
{\cal C}^{\phi(1680)KK^*}_{11}  &=&
(c_1^{\phi_{1680}}+c_2^{\phi_{1680}})
\left\{\frac{}{}f_5(p_{KK^\ast},\beta_{\phi_{1680}},\beta_K,\beta_K^\ast)
\,e_1(p_{KK^\ast},\beta_{\phi_{1680}},\beta_K,\beta_K^\ast) \right.
\nn\\
&&\left.+c_1^{\phi_\Delta}c_2^{\phi_\Delta}
f_6(p_{KK^\ast},\beta_{\phi_{1680}},\beta_K,\beta_{K^\ast},
\beta_{\phi_\Delta})\,e_2(p_{KK^\ast},\beta_{\phi_{1680}},\beta_K,\beta_{K^\ast},
\beta_{\phi_\Delta}) \right.
\nn\\
&&\left.+c_1^{\omega_\Delta}
c_2^{\omega_\Delta}f_6(p_{KK^\ast},\beta_{\phi_{1680}},\beta_K,\beta_{K^\ast},
\beta_{\omega_\Delta})
e_2(p_{KK^\ast},\beta_{\phi_{1680}},\beta_K,\beta_{K^\ast},\beta_{\omega_\Delta})
\frac{}{}\right\}
\label{hfic3po-phi1680-2}
\end{eqnarray}
\begin{eqnarray}
f_5(p,\beta_A,\beta_B,\beta_C)&=&-\frac{\sqrt{2}\,p\, \beta_B^{3/2}
\beta_C^{3/2} \beta_A^{3/2}}{3\left(\beta_C^2 \beta_A^2+\beta_B^2
\left(\beta_C^2+\beta_A^2\right)\right)^{9/2}} \left[\frac{}{}12
\beta_B^6 \beta_C^6 -2 \beta_B^4 \beta_C^4
\left(\beta_B^2+\beta_C^2\right) \beta_A^2  -20 \beta_B^2 \beta_C^2
\left(\beta_B^2+\beta_C^2\right)^2 \beta_A^4\right.
\nn\\
&&\left.-6 \left(\beta_B^2+\beta_C^2\right)^3 \beta_A^6
+p^2\left(\beta_B^2+\beta_C^2\right)^2 \beta_A^2 \left(2 \beta_B^2
\beta_C^2+\left(\beta_B^2+\beta_C^2\right) \beta_A^2\right) \right]
\label{hfic3po-f5}
\end{eqnarray}
\begin{eqnarray}
f_6(p,\beta_A,\beta_B,\beta_C,\beta)&=&-\frac{2 \sqrt{2}\,p\,
\beta_B^{3/2} \beta_C^{3/2} \beta_A^{3/2}} {9\left(\beta_C^2
\beta_A^2+\beta_B^2 \left(\beta_C^2+\beta_A^2\right)+2
\left(\beta_C^2+\beta_A^2\right) \beta^2\right)^{9/2}} \left[2
\left(\beta_C^2 \beta_A^2 +\beta_B^2
\left(\beta_C^2+\beta_A^2\right)+2 \left(\beta_C^2+\beta_A^2\right)
\beta^2\right)\right.
\nn\\
&&\left.\times \left(\beta_B^4 \left(6 \beta_C^4-7 \beta_C^2
\beta_A^2-3 \beta_A^4\right)-\beta_B^2 \left(7 \beta_C^4 \beta_A^2+6
\beta_C^2 \beta_A^4+2 \left(-9 \beta_C^4+7 \beta_C^2 \beta_A^2+6
\beta_A^4\right) \beta^2\right)-3 \right.\right.
\nn\\
&&\left.\left.\times\left(4 \beta_C^2 \beta_A^4 \beta^2+4 \beta_A^4
\beta^4+\beta_C^4 \left(\beta_A^4-4
\beta^4\right)\right)\right)+\beta_A^2 \left(\beta_B^2+\beta_C^2+2
\beta^2\right)^2 \left(\beta_C^2 \beta_A^2+\beta_B^2 \left(2
\beta_C^2+\beta_A^2\right)\right.\right.
\nn\\
&&\left.\left.+2 \left(\beta_C^2+\beta_A^2\right) \beta^2\right)
p^2\right]
 \label{hfic3po-f6}
\end{eqnarray}

$\rm{\underline{\phi(1680)\rightarrow \eta \phi}:}$

\begin{eqnarray}
{\cal C}^{\phi(1680)\eta\phi}_{11} &=& -2 \left\{\frac{}{}\left[2
c_1^\eta c_1^{\phi_{1680}} c_1^{\phi}+c_2^\eta c_2^{\phi_{1680}}
c_2^{\phi}\right]f_5(p_{\eta\phi},\beta_{\phi_{1680}},\beta_\eta,\beta_\phi)
\,e_1(p_{\eta\phi},\beta_{\phi_{1680}},\beta_\eta,\beta_\phi)
\right.
\nn\\
&&\left.+ \left[2 c_1^\eta c_1^{\phi_{1680}} c_1^{\phi}
(c_1^{\phi_\Delta})^2+c_2^\eta c_2^{\phi_{1680}} c_2^{\phi}
(c_2^{\phi_\Delta})^2\right]f_6(p_{\eta\phi},\beta_{\phi_{1680}},
\beta_\eta,\beta_\phi,\beta_{\phi_\Delta})\,e_2(p_{\eta\phi},\beta_{\phi_{1680}},
\beta_\eta,\beta_\phi,\beta_{\phi_\Delta})\right.
\nn\\
&&\left. +\left[2 c_1^\eta c_1^{\phi_{1680}} c_1^{\phi}
(c_1^{\omega_\Delta})^2+c_2^\eta c_2^{\phi_{1680}} c_2^{\phi}
(c_2^{\omega_\Delta})^2\right]f_6(p_{\eta\phi},\beta_{\phi_{1680}},
\beta_\eta,\beta_\phi,\beta_{\omega_\Delta})\,
e_2(p_{\eta\phi},\beta_{\phi_{1680}},
\beta_\eta,\beta_\phi,\beta_{\omega_\Delta})\frac{}{}\right\}\nn\\
\label{hfic3po-phi1680-3}
\end{eqnarray}

$\rm{\underline{\phi(2050)\rightarrow K^+ K^-}:}$

\begin{eqnarray}
{\cal C}^{\phi(2050)KK}_{10} &=&
(c_1^{\phi_{2050}}-c_2^{\phi_{2050}})
\left\{\frac{}{}f_7(p_{KK},\beta_{\phi_{2050}},\beta_K)\,
e_1(p_{KK},\beta_{\phi_{2050}},\beta_K,\beta_K) \right.
\nn\\
&&\left. +c_1^{\eta_\Delta}
c_2^{\eta_\Delta}\,f_{8}(p_{KK},\beta_{\phi_{2050}},
\beta_K,\beta_{\eta_\Delta})\, e_2(p_{KK},\beta_{\phi_{2050}},
\beta_K,\beta_K,\beta_{\eta_\Delta}) \right.
\nn\\
&&\left. +c_1^{\eta_\Delta^\prime}
c_2^{\eta_\Delta^\prime}f_{8}(p_{KK},\beta_{\phi_{2050}},
\beta_K,\beta_{\eta_\Delta^\prime})\,e_2(p_{KK},\beta_{\phi_{2050}},
\beta_K,\beta_K,\beta_{\eta_\Delta^\prime})\frac{}{}\right\}
\label{hfic3po-phi2050-1}
\end{eqnarray}
\begin{eqnarray}
f_7(p,\beta_A,\beta_B)&=& \frac{2 \sqrt{2}\,p\,\beta_A^{3/2}}{3
\sqrt{5}\left(\beta_B^2+2 \beta_A^2\right)^{13/2}} \left[5
\left(\beta_B^2+2 \beta_A^2\right)^2 \left(3 \beta_B^6-17 \beta_B^4
\beta_A^2+16 \beta_B^2 \beta_A^4+12 \beta_A^6\right)-4 \beta_A^2
\left(\beta_B^2+2 \beta_A^2\right) \right.
\nn\\
&&\left. \times\left(-5 \beta_B^4+9 \beta_B^2 \beta_A^2+10
\beta_A^4\right) p^2+4 \beta_A^4 \left(\beta_B^2+\beta_A^2\right)
p^4\right]
\label{hfic3po-f7}
\end{eqnarray}
\begin{eqnarray}
f_{8}(p,\beta_A,\beta_B,\beta)&=&\frac{4 \sqrt{2}\,p\,\beta_A^{3/2}
\beta_B^3\left(\beta_B^2+\beta^2\right) }{9
\sqrt{5}\left(\beta_B^4+2 \beta^2 \beta_A^2+2 \beta_B^2
\left(\beta^2+\beta_A^2\right)\right)^{13/2}} \left[5
\left(\beta_B^4 +2 \beta_B^2
\beta^2-2\beta_A^2\left(\beta_B^2+\beta^2\right) \right) \left(3
\beta_B^4 \left(\beta_B^2+2 \beta^2\right)\right.\right.
\nn\\
&&\left.\left.-11 \beta_B^4
\beta_A^2-6\left(\beta_B^2+\beta^2\right)
\beta_A^4\right)\left(\beta_B^4+2 \beta^2 \beta_A^2+2 \beta_B^2
\left(\beta^2+\beta_A^2\right)\right)^2+4\,
p^2\left(\beta_B^2+\beta^2\right)^2 \beta_A^2 \right.
\nn\\
&&\left.\times  \left(5 \left(\beta_B^6+2 \beta_B^4 \beta^2\right)-9
\beta_B^4 \beta_A^2-10
\left(\beta_B^2+\beta^2\right)\beta_A^4\right) \left(\beta_B^4+2
\beta^2 \beta_A^2+2 \beta_B^2
\left(\beta^2+\beta_A^2\right)\right)\right.
\nn\\
&&\left.+4\, p^4 \left(\beta_B^2+\beta^2\right)^4 \beta_A^4
\left(\beta_B^2+\beta_A^2\right)\right]
\label{hfic3po-f8}
\end{eqnarray}

$\rm{\underline{\phi(2050)\rightarrow K^+ K^{\ast-}}:}$

\begin{eqnarray}
{\cal C}^{\phi(2050)KK^{\ast-}}_{11}&=&
(c_1^{\phi_{2050}}+c_2^{\phi_{2050}})\left\{\frac{}{}f_{9}(p_{KK^\ast},\beta_{\phi_{2050}},
\beta_K,\beta_{K^\ast})\,e_1(p_{KK^\ast},\beta_{\phi_{2050}},
\beta_K,\beta_{K^\ast}) \right.
\nn\\
&&\left.  +c_1^{\phi_\Delta} c_2^{\phi_\Delta}
f_{10}(p_{KK^\ast},\beta_{\phi_{2050}},
\beta_K,\beta_{K^\ast},\beta_{\phi_\Delta})\,
e_2(p_{KK^\ast},\beta_{\phi_{2050}},
\beta_K,\beta_{K^\ast},\beta_{\phi_\Delta})\right.
\nn\\
&&\left. +c_1^{\omega_\Delta}
c_2^{\omega_\Delta}f_{10}(p_{KK^\ast},\beta_{\phi_{2050}},
\beta_K,\beta_{K^\ast},\beta_{\omega_\Delta})\,e_2(p_{KK^\ast},\beta_{\phi_{2050}},
\beta_K,\beta_{K^\ast},\beta_{\omega_\Delta})\frac{}{}\right\}
\label{hfic3po-phi2050-2}
\end{eqnarray}
\begin{eqnarray}
f_{9}(p,\beta_A,\beta_B,\beta_C)&=&\frac{p\,\beta_B^{3/2}
\beta_C^{3/2} \beta_A^{3/2}}{6\sqrt{10}\left(\beta_C^2
\beta_A^2+\beta_B^2 \left(\beta_C^2+\beta_A^2\right)\right)^{13/2} }
 \left[20 \left(6 \beta_B^{10} \beta_C^{10}-5
\beta_B^8 \beta_C^8 \left(\beta_B^2+\beta_C^2\right) \beta_A^2-20
\beta_B^6 \beta_C^6 \right.\right.
\nn\\
&&\left.\left.\times\left(\beta_B^2+\beta_C^2\right)^2\beta_A^4+2
\beta_B^4 \beta_C^4 \left(\beta_B^2+\beta_C^2\right)^3 \beta_A^6 +14
\beta_B^2 \beta_C^2 \left(\beta_B^2+\beta_C^2\right)^4 \beta_A^8 +3
\left(\beta_B^2+\beta_C^2\right)^5 \beta_A^{10}\right)\right.
\nn\\
&&\left.+4\,p^2 \left(\beta_B^2+\beta_C^2\right)^2 \beta_A^2
\left(10 \beta_B^6 \beta_C^6+\beta_B^4
\beta_C^4\left(\beta_B^2+\beta_C^2\right) \beta_A^2-14 \beta_B^2
\beta_C^2 \left(\beta_B^2+\beta_C^2\right)^2 \beta_A^4-5
\right.\right.
\nn\\
&&\left.\left.\times\left(\beta_B^2+\beta_C^2\right)^3
\beta_A^6\right) +p^4\left(\beta_B^2+\beta_C^2\right)^4 \beta_A^4
\left(2 \beta_B^2 \beta_C^2+\left(\beta_B^2+\beta_C^2\right)
\beta_A^2\right) \right]
\label{hfic3po-f9}
\end{eqnarray}
\begin{eqnarray}
f_{10}(p,\beta_A,\beta_B,\beta_C,\beta)&=&\frac{ p\,\beta_B^{3/2}
\beta_C^{3/2} \beta_A^{3/2}}{9\sqrt{10}\left(\beta_C^2
\beta_A^2+\beta_B^2 \left(\beta_C^2+\beta_A^2\right)+2
\left(\beta_C^2+\beta_A^2\right) \beta^2\right)^{13/2}} \left[20
\left(-\beta_C^2 \beta_A^2+\beta_B^2
\left(\beta_C^2-\beta_A^2\right)\right.\right.
\nn\\
&&\left.\left.+2 \left(\beta_C^2-\beta_A^2\right) \beta^2\right)
\left(\beta_C^2 \beta_A^2+\beta_B^2
\left(\beta_C^2+\beta_A^2\right)+2 \left(\beta_C^2+\beta_A^2\right)
\beta^2\right)^2 \left(\beta_B^4 \left(6 \beta_C^4-11 \beta_C^2
\right.\right.\right.
\nn\\
&&\left.\left.\left. \times \beta_A^2-3 \beta_A^4\right)-\beta_B^2
\left(11 \beta_C^4 \beta_A^2+6 \beta_C^2 \beta_A^4+2\beta^2 \left(-9
\beta_C^4+11 \beta_C^2 \beta_A^2 +6 \beta_A^4\right) \right)-3
\left(4 \beta_C^2 \beta_A^4 \beta^2\right.\right.\right.
\nn\\
&&\left.\left.\left.+4 \beta_A^4 \beta^4+\beta_C^4\left(\beta_A^4-4
\beta^4\right)\right)\right)+4 \beta_A^2 \left(\beta_B^2+\beta_C^2+2
\beta^2\right)^2 \left(\beta_C^2 \beta_A^2+\beta_B^2
\left(\beta_C^2+\beta_A^2\right)+2\right.\right.
\nn\\
&&\left.\left.\times \left(\beta_C^2+\beta_A^2\right)
\beta^2\right)\left(\beta_B^4 \left(10 \beta_C^4-9 \beta_C^2
\beta_A^2
 -5 \beta_A^4\right)-\beta_B^2 \left(20
\beta_A^4 \beta^2+\beta_C^4 \left(9 \beta_A^2-30 \beta^2\right)
\right.\right.\right.
\nn\\
&&\left.\left.\left.+2 \beta_C^2 \beta_A^2 \left(5 \beta_A^2+9
\beta^2\right)\right)-5 \left(4 \beta_C^2 \beta_A^4 \beta^2+4
\beta_A^4 \beta^4+\beta_C^4 \left(\beta_A^4-4
\beta^4\right)\right)\right) p^2+\beta_A^4\right.
\nn\\
&&\left.\times \left(\beta_B^2+\beta_C^2+2 \beta^2\right)^4
\left(\beta_C^2 \beta_A^2 +\beta_B^2 \left(2
\beta_C^2+\beta_A^2\right)+2 \left(\beta_C^2+\beta_A^2\right)
\beta^2\right) p^4\right]
 \label{hfic3po-f10}
\end{eqnarray}

$\rm{\underline{\phi(2050)\rightarrow K^{\ast+}K^{\ast-}}:}$

\begin{eqnarray}
{\cal C}^{\phi(2050)K^{\ast+}K^{\ast-}}_{12}&=&
2\,(c_1^{\phi_{2050}}-c_2^{\phi_{2050}})
\left\{\frac{}{}-f_7(p_{K^\ast K^\ast},\beta_{\phi_{2050}},
\beta_{K^\ast})\,e_1(p_{K^\ast K^\ast},\beta_{\phi_{2050}},
\beta_{K^\ast},\beta_{K^\ast}) \right.
\nn\\
&&\left. +2c_1^{\phi_\Delta} c_2^{\phi_\Delta}f_{8}(p_{K^\ast
K^\ast},\beta_{\phi_{2050}},
\beta_{K^\ast},\beta_{\phi_\Delta})\,e_2(p_{K^\ast
K^\ast},\beta_{\phi_{2050}},
\beta_{K^\ast},\beta_{K^\ast},\beta_{\phi_\Delta}) \right.
\nn\\
&&\left.+2c_1^{\omega_\Delta} c_2^{\omega_\Delta}f_{8}(p_{K^\ast
K^\ast},\beta_{\phi_{2050}},
\beta_{K^\ast},\beta_{\omega_\Delta})\,e_2(p_{K^\ast
K^\ast},\beta_{\phi_{2050}},
\beta_{K^\ast},\beta_{K^\ast},\beta_{\omega_\Delta})\frac{}{}\right\}\nn\\
\label{hfic3po-phi2050-3}
\end{eqnarray}

$\rm{\underline{\phi(2050)\rightarrow K^+ K_1^-(1270)}:}$

\begin{eqnarray}
{\cal C}^{\phi(2050)KK_1(1270)}_{01}&=& -(c_1^{\phi_{2050}} +
c_2^{\phi_{2050}})\left\{\frac{}{}f_{11}(p_{KK_1},\beta_{\phi_{2050}}
,\beta_{K},\beta_{K_1})\, e_1(p_{KK_1},\beta_{\phi_{2050}},
\beta_K,\beta_{K_1}) \right.
\nn\\
&&\left. + c_1^{\eta_\Delta}c_2^{\eta_\Delta} f_{12}(p_{K
K_1},\beta_{\phi_{2050}},
\beta_{K},\beta_{K_1},\beta_{\eta_\Delta})\, e_2(p_{K
K_1},\beta_{\phi_{2050}}, \beta_{K},\beta_{K_1},\beta_{\eta_\Delta})
\right.
\nn\\
&&\left.+ c_1^{\eta_\Delta^\prime}
c_2^{\eta_\Delta^\prime}f_{12}(p_{K K_1},\beta_{\phi_{2050}},
\beta_{K},\beta_{K_1},\beta_{\eta^\prime_\Delta})\, e_2(p_{K
K_1},\beta_{\phi_{2050}},
\beta_{K},\beta_{K_1},\beta_{\eta^\prime_\Delta})\frac{}{}\right\}\cos\theta
\nn\\
&& + \frac{(c_1^{\phi_{2050}} -
c_2^{\phi_{2050}})}{\sqrt{2}}\left\{\frac{}{}2\,f_{11}(p_{KK_1},\beta_{\phi_{2050}}
,\beta_{K},\beta_{K_1})\, e_1(p_{KK_1},\beta_{\phi_{2050}},
\beta_K,\beta_{K_1})\right.
\nn\\
&&\left. - c_1^{\phi_\Delta} c_2^{\phi_\Delta}f_{12}(p_{K
K_1},\beta_{\phi_{2050}},
\beta_{K},\beta_{K_1},\beta_{\phi_\Delta})\, e_2(p_{K
K_1},\beta_{\phi_{2050}},
\beta_{K},\beta_{K_1},\beta_{\phi_\Delta})\right.
\nn\\
&&\left. - c_1^{\omega_\Delta} c_2^{\omega_\Delta}f_{12}(p_{K
K_1},\beta_{\phi_{2050}},
\beta_{K},\beta_{K_1},\beta_{\omega_\Delta})\, e_2(p_{K
K_1},\beta_{\phi_{2050}},
\beta_{K},\beta_{K_1},\beta_{\omega_\Delta})\frac{}{}\right\}\sin\theta
\label{hfic3po-phi2050-4.a}
\end{eqnarray}
\begin{eqnarray}
 {\cal C}^{\phi(2050)KK_1(1270)}_{21}&=& -(c_1^{\phi_{2050}}
+
c_2^{\phi_{2050}})\left\{\frac{}{}f_{13}(p_{KK_1},\beta_{\phi_{2050}}
,\beta_{K},\beta_{K_1})\, e_1(p_{KK_1},\beta_{\phi_{2050}},
\beta_K,\beta_{K_1}) \right.
\nn\\
&&\left.+ c_1^{\eta_\Delta}c_2^{\eta_\Delta}f_{14}(p_{K
K_1},\beta_{\phi_{2050}},
\beta_{K},\beta_{K_1},\beta_{\eta_\Delta})\, e_2(p_{K
K_1},\beta_{\phi_{2050}}, \beta_{K},\beta_{K_1},\beta_{\eta_\Delta})
\right.
\nn\\
&&\left.+c_1^{\eta_\Delta^\prime}
c_2^{\eta_\Delta^\prime}f_{14}(p_{K K_1},\beta_{\phi_{2050}},
\beta_{K},\beta_{K_1},\beta_{\eta^\prime_\Delta})\, e_2(p_{K
K_1},\beta_{\phi_{2050}},
\beta_{K},\beta_{K_1},\beta_{\eta^\prime_\Delta})\frac{}{}
\right\}\cos\theta
\nn\\
&& - \frac{(c_1^{\phi_{2050}} -
c_2^{\phi_{2050}})}{\sqrt{2}}\left\{\frac{}{}f_{13}(p_{KK_1},\beta_{\phi_{2050}}
,\beta_{K},\beta_{K_1})\, e_1(p_{KK_1},\beta_{\phi_{2050}},
\beta_K,\beta_{K_1})\right.
\nn\\
&&\left. - 2c_1^{\phi_\Delta} c_2^{\phi_\Delta}f_{14}(p_{K
K_1},\beta_{\phi_{2050}},
\beta_{K},\beta_{K_1},\beta_{\phi_\Delta})\, e_2(p_{K
K_1},\beta_{\phi_{2050}},
\beta_{K},\beta_{K_1},\beta_{\phi_\Delta})\right.
\nn\\
&&\left. - 2c_1^{\omega_\Delta} c_2^{\omega_\Delta}f_{14}(p_{K
K_1},\beta_{\phi_{2050}},
\beta_{K},\beta_{K_1},\beta_{\omega_\Delta})\, e_2(p_{K
K_1},\beta_{\phi_{2050}},
\beta_{K},\beta_{K_1},\beta_{\omega_\Delta})\frac{}{}\right\}\sin\theta
\label{hfic3po-phi2050-4.b}
\end{eqnarray}
\begin{eqnarray}
f_{11}(p,\beta_A,\beta_B,\beta_C)&=&\frac{\beta_B^{7/2}
\beta_C^{5/2} \beta_A^{3/2}}{12 \sqrt{15} \left(\beta_C^2
\beta_A^2+\beta_B^2
\left(\beta_C^2+\beta_A^2\right)\right)^{15/2}}\left[240 \beta_A^2
\left(11 \beta_B^4 \beta_C^4-14 \beta_B^2 \beta_C^2
\left(\beta_B^2+\beta_C^2\right) \beta_A^2\right.\right.
\nn\\
&&\left.\left.+3 \left(\beta_B^2+\beta_C^2\right)^2 \beta_A^4\right)
\left(\beta_C^2 \beta_A^2+\beta_B^2
\left(\beta_C^2+\beta_A^2\right)\right)^3+20 \left(6 \beta_B^6
\beta_C^6-33 \beta_B^4 \beta_C^4 \left(\beta_B^2+\beta_C^2\right)
\beta_A^2\right.\right.
\nn\\
&&\left.\left.+44 \beta_B^2 \beta_C^2
\left(\beta_B^2+\beta_C^2\right)^2
\beta_A^4-\left(\beta_B^2+\beta_C^2\right)^3 \beta_A^6\right)
\left(\beta_C^2 \beta_A^2+\beta_B^2
\left(\beta_C^2+\beta_A^2\right)\right)^2 p^2+4 \right.
\nn\\
&&\left.\times \left(\beta_B^2+\beta_C^2\right)^2 \beta_A^2 \left(10
\beta_B^6 \beta_C^6-7 \beta_B^4 \beta_C^4
\left(\beta_B^2+\beta_C^2\right) \beta_A^2-23 \beta_B^2 \beta_C^2
\left(\beta_B^2+\beta_C^2\right)^2 \beta_A^4\right.\right.
\nn\\
&&\left.\left.-6 \left(\beta_B^2+\beta_C^2\right)^3 \beta_A^6\right)
p^4+\left(\beta_B^2+\beta_C^2\right)^4 \beta_A^4 \left(2 \beta_B^2
\beta_C^2+\left(\beta_B^2+\beta_C^2\right) \beta_A^2\right)
p^6\right]
\label{hfic3po-f11}
\end{eqnarray}
\begin{eqnarray}
f_{12}(p,\beta_A,\beta_B,\beta_C,\beta)&=&-\frac{\beta_B^{3/2}
\beta_C^{5/2} \beta_A^{3/2}}{18 \sqrt{15} \left(\beta_C^2
\left(\beta_B^2+2 \beta^2\right)+\left(\beta_B^2+\beta_C^2+2
\beta^2\right) \beta_A^2\right)^{15/2} }\left[240 \beta_B^2
\beta_A^2 \left(\beta_C^2 \left(\beta_B^2+2
\beta^2\right)\right.\right.
\nn\\
&&\left.\left.+\left(\beta_B^2+\beta_C^2+2 \beta^2\right)
\beta_A^2\right)^3 \left(11 \beta_C^4 \left(\beta_B^2+2
\beta^2\right)^2-14 \beta_C^2 \left(\beta_B^2+2 \beta^2\right)
\left(\beta_B^2+\beta_C^2+2 \beta^2\right) \beta_A^2\right.\right.
\nn\\
&&\left.\left.+3 \left(\beta_B^2+\beta_C^2+2 \beta^2\right)^2
\beta_A^4\right)+20 \left(\beta_C^2 \left(\beta_B^2+2
\beta^2\right)+\left(\beta_B^2+\beta_C^2+2 \beta^2\right)
\beta_A^2\right)^2 \left(6 \beta_C^6
\left(\beta_B^2+\beta^2\right)\right.\right.
\nn\\
&&\left.\left. \times\left(\beta_B^2+2 \beta^2\right)^3-11 \beta_C^4
\left(\beta_B^2+2 \beta^2\right)^2 \left(3 \beta_B^2+2
\beta^2\right) \left(\beta_B^2+\beta_C^2+2 \beta^2\right)
\beta_A^2+2 \beta_C^2\right.\right.
\nn\\
&&\left.\left.\times \left(\beta_B^2+\beta_C^2+2 \beta^2\right)^2
\left(22 \beta_B^4+41 \beta_B^2 \beta^2-6 \beta^4\right)
\beta_A^4-\left(\beta_B^2-22 \beta^2\right)
\left(\beta_B^2+\beta_C^2+2 \beta^2\right)^3 \beta_A^6\right)
p^2\right.
\nn\\
&&\left.+4 \left(\beta_B^2+\beta_C^2+2 \beta^2\right)^2 \beta_A^2
\left(10 \beta_C^6 \left(\beta_B^2+\beta^2\right) \left(\beta_B^2+2
\beta^2\right)^3-\beta_C^4 \left(7 \beta_B^2-2 \beta^2\right)
\left(\beta_B^2+2 \beta^2\right)^2 \right.\right.
\nn\\
&&\left.\left.\times\left(\beta_B^2+\beta_C^2+2 \beta^2\right)
\beta_A^2-\beta_C^2 \left(\beta_B^2+\beta_C^2+2 \beta^2\right)^2
\left(23 \beta_B^4+72 \beta_B^2 \beta^2+52 \beta^4\right)
\beta_A^4-6 \right.\right.
\nn\\
&&\left.\left.\times\left(\beta_B^2+\beta_C^2+2 \beta^2\right)^3
\left(\beta_B^2+3 \beta^2\right) \beta_A^6\right)
p^4+\left(\beta_B^2+2 \beta^2\right) \left(\beta_B^2+\beta_C^2+2
\beta^2\right)^4 \beta_A^4 \left(2 \beta_C^2\right.\right.
\nn\\
&&\left.\left.\times
\left(\beta_B^2+\beta^2\right)+\left(\beta_B^2+\beta_C^2+2
\beta^2\right) \beta_A^2\right) p^6\right]
\label{hfic3po-f12}
\end{eqnarray}
\begin{eqnarray}
f_{13}(p,\beta_A,\beta_B,\beta_C)&=&-\frac{\beta_B^{7/2}
\beta_C^{5/2} \beta_A^{3/2} p^2}{60 \sqrt{3} \left(\beta_C^2
\beta_A^2+\beta_B^2 \left(\beta_C^2+\beta_A^2\right)\right)^{15/2}}
\left[120 \beta_B^{10} \beta_C^{10}-420 \beta_B^8 \beta_C^8
\left(\beta_B^2+\beta_C^2\right) \beta_A^2-752 \beta_B^6
\beta_C^6\right.
\nn\\
&&\left.\times \left(\beta_B^2+\beta_C^2\right)^2 \beta_A^4+456
\beta_B^4 \beta_C^4 \left(\beta_B^2+\beta_C^2\right)^3 \beta_A^6+888
\beta_B^2 \beta_C^2 \left(\beta_B^2+\beta_C^2\right)^4
\beta_A^8+220\right.
\nn\\
&&\left.\times \left(\beta_B^2+\beta_C^2\right)^5 \beta_A^{10}+4
\left(\beta_B^2+\beta_C^2\right)^2 \beta_A^2 \left(10 \beta_B^6
\beta_C^6-7 \beta_B^4 \beta_C^4 \left(\beta_B^2+\beta_C^2\right)
\beta_A^2-26 \beta_B^2 \beta_C^2\right.\right.
\nn\\
&&\left.\left.\times \left(\beta_B^2+\beta_C^2\right)^2 \beta_A^4-9
\left(\beta_B^2+\beta_C^2\right)^3 \beta_A^6\right)
p^2+\left(\beta_B^2+\beta_C^2\right)^4 \beta_A^4 \left(2 \beta_B^2
\beta_C^2+\left(\beta_B^2+\beta_C^2\right) \beta_A^2\right)
p^4\right]\nn\\
\label{hfic3po-f13}
\end{eqnarray}
\begin{eqnarray}
f_{14}(p,\beta_A,\beta_B,\beta_C,\beta)&=&\frac{\beta_B^{3/2}
\beta_C^{5/2} \left(\beta_B^2+2 \beta^2\right) \beta_A^{3/2}p^2}{90
\sqrt{3} \left(\beta_C^2 \left(\beta_B^2+2
\beta^2\right)+\left(\beta_B^2+\beta_C^2+2 \beta^2\right)
\beta_A^2\right)^{15/2} }\left[4 \left(30 \beta_C^{10}
\left(\beta_B^2+\beta^2\right) \left(\beta_B^2+2
\beta^2\right)^4\right.\right.
\nn\\
&&\left.\left.-5 \beta_C^8 \left(\beta_B^2+2 \beta^2\right)^3
\left(\beta_B^2+\beta_C^2+2 \beta^2\right) \left(21 \beta_B^2+10
\beta^2\right) \beta_A^2-4 \beta_A^4\beta_C^6 \left(\beta_B^2+2
\beta^2\right)^2 \right.\right.
\nn\\
&&\left.\left.\times\left(\beta_B^2+\beta_C^2+2 \beta^2\right)^2
\left(47 \beta_B^2+55 \beta^2\right) +6 \beta_C^4
\left(\beta_B^2+\beta_C^2+2 \beta^2\right)^3 \left(19 \beta_B^4+28
\beta_B^2 \beta^2-20 \beta^4\right)\right.\right.
\nn\\
&&\left.\left.\times \beta_A^6+2 \beta_C^2
\left(\beta_B^2+\beta_C^2+2 \beta^2\right)^4 \left(111 \beta_B^2+95
\beta^2\right) \beta_A^8+55 \left(\beta_B^2+\beta_C^2+2
\beta^2\right)^5 \beta_A^{10}\right)+4 \right.
\nn\\
&&\left.\times\left(\beta_B^2+\beta_C^2+2 \beta^2\right)^2 \beta_A^2
\left(10 \beta_C^6 \left(\beta_B^2+\beta^2\right) \left(\beta_B^2+2
\beta^2\right)^2-\beta_C^4 \left(\beta_B^2+\beta_C^2+2
\beta^2\right)\right.\right.
\nn\\
&&\left.\left.\times \left(7 \beta_B^4+12 \beta_B^2 \beta^2-4
\beta^4\right) \beta_A^2-26 \beta_C^2 \left(\beta_B^2+\beta^2\right)
\left(\beta_B^2+\beta_C^2+2 \beta^2\right)^2
\beta_A^4-9\beta_A^6\right.\right.
\nn\\
&&\left.\left.\times \left(\beta_B^2+\beta_C^2+2 \beta^2\right)^3
\right) p^2+\left(\beta_B^2+\beta_C^2+2 \beta^2\right)^4 \beta_A^4
\left(2 \beta_C^2
\left(\beta_B^2+\beta^2\right)+\left(\beta_B^2+\beta_C^2+2
\beta^2\right) \beta_A^2\right) p^4\right]\nn\\
\label{hfic3po-f14}
\end{eqnarray}

$\rm{\underline{\phi(2050)\rightarrow K^+ K_1^-(1400)}:}$

\begin{eqnarray}
{\cal C}^{\phi(2050)KK_1(1400)}_{01}&=& (c_1^{\phi_{2050}} +
c_2^{\phi_{2050}})\left\{\frac{}{}f_{11}(p_{KK_1},\beta_{\phi_{2050}}
,\beta_{K},\beta_{K_1})\, e_1(p_{KK_1},\beta_{\phi_{2050}},
\beta_K,\beta_{K_1}) \right.
\nn\\
&&\left. + c_1^{\eta_\Delta}c_2^{\eta_\Delta} f_{12}(p_{K
K_1},\beta_{\phi_{2050}},
\beta_{K},\beta_{K_1},\beta_{\eta_\Delta})\, e_2(p_{K
K_1},\beta_{\phi_{2050}}, \beta_{K},\beta_{K_1},\beta_{\eta_\Delta})
\right.
\nn\\
&&\left.+ c_1^{\eta_\Delta^\prime}
c_2^{\eta_\Delta^\prime}f_{12}(p_{K K_1},\beta_{\phi_{2050}},
\beta_{K},\beta_{K_1},\beta_{\eta^\prime_\Delta})\, e_2(p_{K
K_1},\beta_{\phi_{2050}},
\beta_{K},\beta_{K_1},\beta_{\eta^\prime_\Delta})\frac{}{}\right\}\sin\theta
\nn\\
&& + \frac{(c_1^{\phi_{2050}} -
c_2^{\phi_{2050}})}{\sqrt{2}}\left\{\frac{}{}2\,f_{11}(p_{KK_1},\beta_{\phi_{2050}}
,\beta_{K},\beta_{K_1})\, e_1(p_{KK_1},\beta_{\phi_{2050}},
\beta_K,\beta_{K_1})\right.
\nn\\
&&\left. - c_1^{\phi_\Delta} c_2^{\phi_\Delta}f_{12}(p_{K
K_1},\beta_{\phi_{2050}},
\beta_{K},\beta_{K_1},\beta_{\phi_\Delta})\, e_2(p_{K
K_1},\beta_{\phi_{2050}},
\beta_{K},\beta_{K_1},\beta_{\phi_\Delta})\right.
\nn\\
&&\left. - c_1^{\omega_\Delta} c_2^{\omega_\Delta}f_{12}(p_{K
K_1},\beta_{\phi_{2050}},
\beta_{K},\beta_{K_1},\beta_{\omega_\Delta})\, e_2(p_{K
K_1},\beta_{\phi_{2050}},
\beta_{K},\beta_{K_1},\beta_{\omega_\Delta})\frac{}{}\right\}\cos\theta
\label{hfic3po-phi2050-5.a}
\end{eqnarray}
\begin{eqnarray}
{\cal C}^{\phi(2050)KK_1(1400)}_{21}&=& (c_1^{\phi_{2050}} +
c_2^{\phi_{2050}})\left\{\frac{}{}f_{13}(p_{KK_1},\beta_{\phi_{2050}}
,\beta_{K},\beta_{K_1})\, e_1(p_{KK_1},\beta_{\phi_{2050}},
\beta_K,\beta_{K_1}) \right.
\nn\\
&&\left.+ c_1^{\eta_\Delta}c_2^{\eta_\Delta}f_{14}(p_{K
K_1},\beta_{\phi_{2050}},
\beta_{K},\beta_{K_1},\beta_{\eta_\Delta})\, e_2(p_{K
K_1},\beta_{\phi_{2050}}, \beta_{K},\beta_{K_1},\beta_{\eta_\Delta})
\right.
\nn\\
&&\left.+c_1^{\eta_\Delta^\prime}
c_2^{\eta_\Delta^\prime}f_{14}(p_{K K_1},\beta_{\phi_{2050}},
\beta_{K},\beta_{K_1},\beta_{\eta^\prime_\Delta})\, e_2(p_{K
K_1},\beta_{\phi_{2050}},
\beta_{K},\beta_{K_1},\beta_{\eta^\prime_\Delta})
\frac{}{}\right\}\sin\theta
\nn\\
&& - \frac{(c_1^{\phi_{2050}} -
c_2^{\phi_{2050}})}{\sqrt{2}}\left\{\frac{}{}f_{13}(p_{KK_1},\beta_{\phi_{2050}}
,\beta_{K},\beta_{K_1})\, e_1(p_{KK_1},\beta_{\phi_{2050}},
\beta_K,\beta_{K_1})\right.
\nn\\
&&\left. - 2c_1^{\phi_\Delta} c_2^{\phi_\Delta}f_{14}(p_{K
K_1},\beta_{\phi_{2050}},
\beta_{K},\beta_{K_1},\beta_{\phi_\Delta})\, e_2(p_{K
K_1},\beta_{\phi_{2050}},
\beta_{K},\beta_{K_1},\beta_{\phi_\Delta})\right.
\nn\\
&&\left. - 2c_1^{\omega_\Delta} c_2^{\omega_\Delta}f_{14}(p_{K
K_1},\beta_{\phi_{2050}},
\beta_{K},\beta_{K_1},\beta_{\omega_\Delta})\, e_2(p_{K
K_1},\beta_{\phi_{2050}},
\beta_{K},\beta_{K_1},\beta_{\omega_\Delta})\frac{}{}\right\}\cos\theta)
\label{hfic3po-phi2050-5.b}
\end{eqnarray}

$\rm{\underline{\phi(2050)\rightarrow K^+ K_0^{\ast -}(1430)}:}$

\begin{eqnarray}
{\cal C}^{\phi(2050)KK_0^\ast(1430)}_{20}&=&
(c_1^{\phi_{2050}}-c_2^{\phi_{2050}})\left\{\frac{}{}c_1^{\phi_\Delta}
c_2^{\phi_\Delta}
f_{14}(p_{KK_0^\ast},\beta_{\phi_{2050}},\beta_K,\beta_{K_0^{\ast}},\beta_{\phi_\Delta})\,
e_2(p_{KK_0^\ast},\beta_{\phi_{2050}},\beta_K,\beta_{K_0^{\ast}},\beta_{\phi_\Delta})
\right.
\nn\\
&&\left. +c_1^{\omega_\Delta} c_2^{\omega_\Delta}
f_{14}(p_{KK_0^\ast},\beta_{\phi_{2050}},\beta_K,\beta_{K_0^{\ast}},\beta_{\omega_\Delta})
\,e_2(p_{KK_0^\ast},\beta_{\phi_{2050}},\beta_K,\beta_{K_0^{\ast}},\beta_{\omega_\Delta})\frac{}{}
\right\} \label{hfic3po-phi2050-6}
\end{eqnarray}

$\rm{\underline{\phi(2050)\rightarrow K^+ K_2^{\ast -}(1430)}:}$

\begin{eqnarray}
{\cal C}^{\phi(2050)KK_2\ast(1430)}_{22}&=&
\sqrt{3}\,(c_1^{\phi_{2050}}-c_2^{\phi_{2050}})
\left\{\frac{}{}f_{13}(p_{KK_2^\ast}
,\beta_{\phi_{2050}},\beta_K,\beta_{K_2^{\ast}})\, e_1(p_{KK_2^\ast}
,\beta_{\phi_{2050}},\beta_K,\beta_{K_2^{\ast}}) \right.
\nn\\
&&\left. +c_1^{\phi_\Delta} c_2^{\phi_\Delta}f_{14}(p_{KK_2^\ast}
,\beta_{\phi_{2050}},\beta_K,\beta_{K_2^{\ast}},\beta_{\phi_\Delta})
\,e_2(p_{KK_2^\ast},\beta_{\phi_{2050}},\beta_K,\beta_{K_2^{\ast}},\beta_{\phi_\Delta})
\right.
\nn\\
&&\left.
 +c_1^{\omega_\Delta} c_2^{\omega_\Delta}
f_{14}(p_{KK_2^\ast}
,\beta_{\phi_{2050}},\beta_K,\beta_{K_2^{\ast}},\beta_{\omega_\Delta})
\, e_2(p_{KK_2^\ast},\beta_{\phi_{2050}},\beta_K,\beta_{K_2^{\ast}},
\beta_{\omega_\Delta}) \frac{}{}\right\}
\label{hfic3po-phi2050-7}
\end{eqnarray}

$\rm{\underline{\phi(2050)\rightarrow K^+ K^{\ast -}(1410)}:}$

\begin{eqnarray}
{\cal C}^{\phi(2050)KK^\ast(1410)}_{11} &=&
(c_1^{\phi_{2050}}+c_2^{\phi_{2050}})\left\{\frac{}{}f_{15}(p_{KK^\ast}
,\beta_{\phi_{2050}} ,\beta_K,\beta_{K^\ast})\, e_1(p_{KK^\ast}
,\beta_{\phi_{2050}},\beta_K,\beta_{K^{\ast}}) \right.
\nn\\
&&\left. + c_1^{\phi_\Delta}c_2^{\phi_\Delta}\,f_{16}(p_{KK^\ast}
,\beta_{\phi_{2050}},\beta_K, \beta_{K^{\ast}},\beta_{\phi_\Delta})
\, e_2(p_{KK^\ast},\beta_{\phi_{2050}},\beta_K,\beta_{K^{\ast}},
\beta_{\phi_\Delta}) \right.
\nn\\
&&\left. +
c_1^{\omega_\Delta}c_2^{\omega_\Delta}\,f_{16}(p_{KK^\ast}
,\beta_{\phi_{2050}},\beta_K,
\beta_{K^{\ast}},\beta_{\omega_\Delta})\,e_2(p_{KK^\ast},\beta_{\phi_{2050}}
,\beta_K,\beta_{K^{\ast}}, \beta_{\omega_\Delta})\frac{}{}\right\}
\label{hfic3po-phi2050-8}
\end{eqnarray}
\begin{eqnarray}
f_{15}(p,\beta_A,\beta_B,\beta_C)&=&
\frac{\beta_A^{3/2} \beta_B^{3/2} \beta_C^{3/2} p} {24 \sqrt{15}
\left(\beta_B^2 \beta_C^2+\beta_A^2
\left(\beta_B^2+\beta_C^2\right)\right)^{17/2}}\left[40
\left(\beta_B^2 \beta_C^2 +\beta_A^2
\left(\beta_B^2+\beta_C^2\right)\right)^3 \left(18 \beta_B^8
 \beta_C^8+\beta_A^4 \beta_B^4 \beta_C^4 \right.\right.
\nn\\
&&\left.\left.\times\left(275 \beta_B^2-27 \beta_C^2\right)
\left(\beta_B^2 +\beta_C^2\right)-11 \beta_A^6 \beta_B^2 \beta_C^2
\left(\beta_B^2 -3 \beta_C^2\right)
\left(\beta_B^2+\beta_C^2\right)^2-9 \beta_A^8
\left(\beta_B^2-\beta_C^2\right) \right.\right.
\nn\\
&&\left.\left. \times\left(\beta_B^2+\beta_C^2\right)^3-11 \beta_A^2
\beta_B^6 \beta_C^6 \left(19 \beta_B^2+3 \beta_C^2\right)\right)-4
\left(\beta_B^2 \beta_C^2 +\beta_A^2
\left(\beta_B^2+\beta_C^2\right)\right)^2 \left(30 \beta_B^{10}
\beta_C^8\right.\right.
\nn\\
&&\left.\left. +2 \beta_A^4 \beta_B^4 \beta_C^4 \left(265
\beta_B^2-3 \beta_C^2\right) \left(\beta_B^2+\beta_C^2\right)^2+30
\beta_A^8 \left(-\beta_B^2+\beta_C^2\right)
\left(\beta_B^2+\beta_C^2\right)^4 -5 \beta_A^2 \beta_B^6 \beta_C^6
\right.\right.
\nn\\
&&\left.\left.\times\left(\beta_B^2+\beta_C^2\right) \left(61
\beta_B^2+12 \beta_C^2\right)+\beta_A^6 \beta_B^2 \beta_C^2
\left(\beta_B^2 +\beta_C^2\right)^3 \left(79 \beta_B^2+84
\beta_C^2\right)\right) p^2 -2 \beta_A^2
\left(\beta_B^2+\beta_C^2\right)^2 \right.
\nn\\
&&\left.\times\left(20 \beta_B^{10} \beta_C^8+3 \beta_A^8
\left(\beta_B^2 -\beta_C^2\right)
\left(\beta_B^2+\beta_C^2\right)^4-15 \beta_A^4 \beta_B^4 \beta_C^4
\left(\beta_B^2+\beta_C^2\right)^2 \left(5
\beta_B^2+\beta_C^2\right) -4 \beta_A^6 \beta_B^2 \beta_C^2
\right.\right.
\nn\\
&&\left.\left.\times\left(\beta_B^2+\beta_C^2\right)^3 \left(4
\beta_B^2+3 \beta_C^2\right)-6 \beta_A^2 \beta_B^6 \beta_C^6 \left(6
\beta_B^4+7 \beta_B^2 \beta_C^2+\beta_C^4\right)\right)
p^4-\beta_A^4 \beta_B^4 \beta_C^2 \left(\beta_B^2+\beta_C^2\right)^4
\right.
\nn\\
&&\left.\times\left(2 \beta_B^2 \beta_C^2+\beta_A^2 \left(\beta_B^2
+\beta_C^2\right)\right) p^6\right]
\label{hfic3po-f15}
\end{eqnarray}
\begin{eqnarray}
f_{16}(p,\beta_A,\beta_B,\beta_C,\beta)&=&-\frac{\beta_A^{3/2}
\beta_B^{3/2} \beta_C^{3/2} p}{36 \sqrt{15} \left(\beta_C^2
\left(\beta_B^2+2 \beta^2\right)+\beta_A^2
\left(\beta_B^2+\beta_C^2+2 \beta^2\right)\right)^{17/2}}\left[40
\left(\beta_C^2 \left(\beta_B^2+2 \beta^2\right)+\beta_A^2
\right.\right.
\nn\\
&&\left.\left. \times\left(\beta_B^2+\beta_C^2+2
\beta^2\right)\right)^3 \left(18 \beta_C^8
\left(\beta_B^2+\beta^2\right) \left(\beta_B^2+2 \beta^2\right)^3-9
\beta_A^8 \left(\beta_B^2-\beta_C^2+2 \beta^2\right) \right.\right.
\nn\\
&&\left.\left.\times\left(\beta_B^2+\beta_C^2+2 \beta^2\right)^3+11
\beta_A^6 \beta_C^2 \left(\beta_B^2+\beta_C^2+2 \beta^2\right)^2
\left(-\beta_B^4+24 \beta^4+\beta_B^2 \left(3 \beta_C^2+10
\beta^2\right)\right)\right.\right.
\nn\\
&&\left.\left.-11 \beta_A^2 \beta_C^6 \left(\beta_B^2+2
\beta^2\right)^2 \left(19 \beta_B^4+24 \beta^4+\beta_B^2 \left(3
\beta_C^2+50 \beta^2\right)\right)+\beta_A^4 \beta_C^4
\left(\beta_B^2+2 \beta^2\right) \right.\right.
\nn\\
&&\left.\left.\times\left(\beta_B^2+\beta_C^2+2 \beta^2\right)
\left(275 \beta_B^4-36 \beta_C^2 \beta^2+\beta_B^2 \left(-27
\beta_C^2+550 \beta^2\right)\right)\right)-4 \left(\beta_C^2
\left(\beta_B^2+2 \beta^2\right)+\beta_A^2 \right.\right.
\nn\\
&&\left.\left.\times\left(\beta_B^2+\beta_C^2+2
\beta^2\right)\right)^2 \left(30 \beta_C^8
\left(\beta_B^2+\beta^2\right) \left(\beta_B^2+2 \beta^2\right)^4-30
\beta_A^8 \left(\beta_B^2-\beta_C^2+2 \beta^2\right) \right.\right.
\nn\\
&&\left.\left.\times\left(\beta_B^2+\beta_C^2+2 \beta^2\right)^4-5
\beta_A^2 \beta_C^6 \left(\beta_B^2+2 \beta^2\right)^2
\left(\beta_B^2+\beta_C^2+2 \beta^2\right) \left(61 \beta_B^4+4
\beta^2 \left(3 \beta_C^2+25 \beta^2\right)\right.\right.\right.
\nn\\
&&\left.\left.\left.+4 \beta_B^2 \left(3 \beta_C^2+43
\beta^2\right)\right)+\beta_A^6 \beta_C^2
\left(\beta_B^2+\beta_C^2+2 \beta^2\right)^3 \left(79 \beta_B^4+60
\beta_C^2 \beta^2+820 \beta^4+\beta_B^2 \right.\right.\right.
\nn\\
&&\left.\left.\left.\times\left(84 \beta_C^2+568
\beta^2\right)\right)+2 \beta_A^4 \beta_C^4 \left(\beta_B^2+2
\beta^2\right) \left(\beta_B^2+\beta_C^2+2 \beta^2\right)^2
\left(265 \beta_B^4-30 \beta_C^2 \beta^2+190 \beta^4
\right.\right.\right.
\nn\\
&&\left.\left.\left.+\beta_B^2\left(-3 \beta_C^2+625
\beta^2\right)\right)\right) p^2-2 \beta_A^2
\left(\beta_B^2+\beta_C^2+2 \beta^2\right)^2 \left(\beta_C^2
\left(\beta_B^2+2 \beta^2\right)+\beta_A^2\right.\right.
\nn\\
&&\left.\left.\times \left(\beta_B^2+\beta_C^2+2
\beta^2\right)\right)\left(20 \beta_C^6
\left(\beta_B^2+\beta^2\right) \left(\beta_B^2+2 \beta^2\right)^3+3
\beta_A^6 \left(\beta_B^2-\beta_C^2+2 \beta^2\right)\right.\right.
\nn\\
&&\left.\left.\times \left(\beta_B^2+\beta_C^2+2 \beta^2\right)^3-2
\beta_A^2 \beta_C^4 \left(\beta_B^2+2 \beta^2\right)
\left(\beta_B^2+\beta_C^2+2 \beta^2\right) \left(28 \beta_B^4+3
\beta_C^2 \beta^2+38 \beta^4\right.\right.\right.
\nn\\
&&\left.\left.\left.+3 \beta_B^2 \left(\beta_C^2+25
\beta^2\right)\right)-\beta_A^4 \beta_C^2
\left(\beta_B^2+\beta_C^2+2 \beta^2\right)^2 \left(19 \beta_B^4+9
\beta_B^2 \left(\beta_C^2+10 \beta^2\right)+4 \beta^2
\right.\right.\right.
\nn\\
&&\left.\left.\left.\times\left(3 \beta_C^2+26
\beta^2\right)\right)\right) p^4-\beta_A^4 \beta_C^2
\left(\beta_B^2+2 \beta^2\right)^2 \left(\beta_B^2+\beta_C^2+2
\beta^2\right)^4 \left(2 \beta_C^2
\left(\beta_B^2+\beta^2\right)+\beta_A^2 \right.\right.
\nn\\
&&\left.\left.\left(\beta_B^2+\beta_C^2+2 \beta^2\right)\right)
p^6\right] \label{hfic3po-f16}
\end{eqnarray}

$\rm{\underline{\phi(2050)\rightarrow K^+ K^{-}(1460)}:}$

\begin{eqnarray}
{\cal
C}^{\phi(2050)KK(1460)}_{10}&=&-(c_1^{\phi_{2050}}-c_2^{\phi_{2050}})\left\{\frac{}{}f_{15}(p_{KK_{1460}}
,\beta_{\phi_{2050}} ,\beta_K,\beta_{K_{1460}})\, e_1(p_{KK_{1460}}
,\beta_{\phi_{2050}},\beta_K,\beta_{K_{1460}}) \right.
\nn\\
&&\left. + c_1^{\eta_\Delta}c_2^{\eta_\Delta}\,f_{16}(p_{KK_{1460}}
,\beta_{\phi_{2050}},\beta_K, \beta_{K_{1460}},\beta_{\eta_\Delta})
\, e_2(p_{KK_{1460}},\beta_{\phi_{2050}},\beta_K,\beta_{K_{1460}},
\beta_{\eta_\Delta}) \right.
\nn\\
&&\left. +
c_1^{\eta^\prime_\Delta}c_2^{\eta^\prime_\Delta}\,f_{16}(p_{KK_{1460}}
,\beta_{\phi_{2050}},\beta_K,
\beta_{K_{1460}},\beta_{\eta^\prime_\Delta})\,e_2(p_{KK_{1460}},\beta_{\phi_{2050}}
,\beta_K,\beta_{K_{1460}},
\beta_{\eta^\prime_\Delta})\frac{}{}\right\}
\label{hfic3po-phi2050-9}
\end{eqnarray}

$\rm{\underline{\phi(2050)\rightarrow \eta \phi}:}$

\begin{eqnarray}
{\cal C}^{\phi(2050)\eta\phi}_{11} &=& 2\left\{ \left(2 c_1^\eta
c_1^{\phi} c_1^{\phi_{2050}}+c_2^\eta c_2^{\phi}
c_2^{\phi_{2050}}\right)\,f_9(p_{\eta\phi},\beta_{\phi_{2050}},\beta_\eta,
\beta_\phi) \, e_1(p_{\eta\phi}
,\beta_{\phi_{2050}},\beta_\eta,\beta_\phi) \right.
\nn\\
&&\left. - \left(2 c_1^\eta c_1^{\phi} c_1^{\phi_{2050}}
(c_1^{\phi_\Delta})^2+c_2^\eta c_2^{\phi} c_2^{\phi_{2050}}
(c_2^{\phi_\Delta})^2\right)f_{10}(p_{\eta\phi},\beta_{\phi_{2050}},\beta_\eta,\beta_\phi,
\beta_{\phi_\Delta}) \, e_2(p_{\eta\phi},\beta_{\phi_{2050}}
,\beta_\eta,\beta_\phi, \beta_{\phi_\Delta}) \right.
\nn\\
&&\left. - \left(2 c_1^\eta c_1^{\phi} c_1^{\phi_{2050}}
(c_1^{\omega_\Delta})^2+c_2^\eta c_2^{\phi} c_2^{\phi_{2050}}
(c_2^{\omega_\Delta})^2\right)f_{10}(p_{\eta\phi},\beta_{\phi_{2050}},
\beta_\eta,\beta_\phi, \beta_{\omega_\Delta})\,
e_2(p_{\eta\phi},\beta_{\phi_{2050}} ,\beta_\eta,\beta_\phi,
\beta_{\omega_\Delta}) \right\}\nn\\
\label{hfic3po-phi2050-10}
\end{eqnarray}

$\rm{\underline{\phi(2050)\rightarrow \eta^\prime \phi}:}$

\begin{eqnarray}
{\cal C}^{\phi(2050)\eta^\prime\phi}_{11} &=&  2\left\{ \left(2
c_1^{\eta^\prime} c_1^{\phi} c_1^{\phi_{2050}}+c_2^{\eta^\prime}
c_2^{\phi} c_2^{\phi_{2050}}\right)\,f_9(p_{\eta^\prime\phi}
,\beta_{\phi_{2050}},\beta_{\eta^\prime}, \beta_\phi) \,
e_1(p_{\eta^\prime\phi}
,\beta_{\phi_{2050}},\beta_{\eta^\prime},\beta_\phi) \right.
\nn\\
&&\left. - \left(2 c_1^{\eta^\prime} c_1^{\phi} c_1^{\phi_{2050}}
(c_1^{\phi_\Delta})^2+c_2^{\eta^\prime} c_2^{\phi} c_2^{\phi_{2050}}
(c_2^{\phi_\Delta})^2\right)f_{10}(p_{\eta^\prime\phi}
,\beta_{\phi_{2050}},\beta_{\eta^\prime},\beta_\phi,
\beta_{\phi_\Delta})\, e_2(p_{\eta^\prime\phi},\beta_{\phi_{2050}}
,\beta_{\eta^\prime},\beta_\phi, \beta_{\phi_\Delta}) \right.
\nn\\
&&\left. - \left(2 c_1^{\eta^\prime} c_1^{\phi} c_1^{\phi_{2050}}
(c_1^{\omega_\Delta})^2+c_2^{\eta^\prime} c_2^{\phi}
c_2^{\phi_{2050}}
(c_2^{\omega_\Delta})^2\right)f_{10}(p_{\eta^\prime\phi}
,\beta_{\phi_{2050}},\beta_{\eta^\prime},\beta_\phi,
\beta_{\omega_\Delta}) \,
e_2(p_{\eta^\prime\phi},\beta_{\phi_{2050}}
,\beta_{\eta^\prime},\beta_\phi, \beta_{\omega_\Delta})
\right\} \nn\\
\label{hfic3po-phi2050-11}
\end{eqnarray}

$\rm{\underline{\phi(2050)\rightarrow \eta h_1(1380)}:}$

\begin{eqnarray}
{\cal C}^{\phi(2050)\eta h_1}_{01}&=& 2\left\{ -(2 c_1^{h_1}
c_1^\eta c_1^{\phi_{2050}}+c_2^{h_1} c_2^\eta c_2^{\phi_{2050}})
f_{11}(p_{\eta h_1},\beta_{\phi_{2050}} ,\beta_{\eta},\beta_{h_1})\,
e_1(p_{\eta h_1},\beta_{\phi_{2050}} ,\beta_{\eta},\beta_{h_1})
\right.
\nn\\
&&\left. + \left(2 c_1^{h_1} c_1^\eta
(c_1^{\eta_\Delta})^2c_1^{\phi_{2050}}+c_2^{h_1} c_2^\eta
(c_2^{\eta_\Delta})^2 c_2^{\phi_{2050}}\right) f_{12}(p_{\eta
h_1},\beta_{\phi_{2050}} ,\beta_{\eta},\beta_{h_1},
\beta_{\eta_\Delta})\, e_2(p_{\eta h_1},\beta_{\phi_{2050}}
,\beta_{\eta},\beta_{h_1}, \beta_{\eta_\Delta}) \right.
\nn\\
&&\left.+ \left(2 c_1^{h_1} c_1^\eta
(c_1^{\eta_\Delta^\prime})^2c_1^{\phi_{2050}}+c_2^{h_1} c_2^\eta
(c_2^{\eta_\Delta^\prime})^2 c_2^{\phi_{2050}}\right) f_{12}(p_{\eta
h_1},\beta_{\phi_{2050}} ,\beta_{\eta},\beta_{h_1},
\beta_{\eta_\Delta^\prime})\, e_2(p_{\eta h_1},\beta_{\phi_{2050}}
,\beta_{\eta},\beta_{h_1}, \beta_{\eta_\Delta^\prime}) \right\}\nn\\
\label{hfic3po-phi2050-12.a}
\end{eqnarray}
\begin{eqnarray}
{\cal C}^{\phi(2050)\eta h_1}_{21}&=& 2\left\{ -(2 c_1^{h_1}
c_1^\eta c_1^{\phi_{2050}}+c_2^{h_1} c_2^\eta c_2^{\phi_{2050}})
f_{13}(p_{\eta h_1},\beta_{\phi_{2050}} ,\beta_{\eta},\beta_{h_1})\,
e_1(p_{\eta h_1},\beta_{\phi_{2050}} ,\beta_{\eta},\beta_{h_1})
\right.
\nn\\
&&\left. + \left(2 c_1^{h_1} c_1^\eta
(c_1^{\eta_\Delta})^2c_1^{\phi_{2050}}+c_2^{h_1} c_2^\eta
(c_2^{\eta_\Delta})^2 c_2^{\phi_{2050}}\right) f_{14}(p_{\eta
h_1},\beta_{\phi_{2050}} ,\beta_{\eta},\beta_{h_1},
\beta_{\eta_\Delta})\, e_2(p_{\eta h_1},\beta_{\phi_{2050}}
,\beta_{\eta},\beta_{h_1}, \beta_{\eta_\Delta}) \right.
\nn\\
&&\left.+ \left(2 c_1^{h_1} c_1^\eta
(c_1^{\eta_\Delta^\prime})^2c_1^{\phi_{2050}}+c_2^{h_1} c_2^\eta
(c_2^{\eta_\Delta^\prime})^2 c_2^{\phi_{2050}}\right) f_{14}(p_{\eta
h_1},\beta_{\phi_{2050}} ,\beta_{\eta},\beta_{h_1},
\beta_{\eta_\Delta^\prime})\, e_2(p_{\eta h_1},\beta_{\phi_{2050}}
,\beta_{\eta},\beta_{h_1}, \beta_{\eta_\Delta^\prime}) \right\}\nn\\
\label{hfic3po-phi2050-12.b}
\end{eqnarray}





$\rm{\underline{\phi_1(1850)\rightarrow K^{\ast+} K^{\ast-}}:}$

\begin{eqnarray}
{\cal C}^{\phi_1 K^\ast K^\ast}_{12}&=&
(c_1^{\phi_1}-c_2^{\phi_1})\left\{\frac{}{}f_{17}(p_{K^\ast
K^\ast},\beta_{\phi_1} ,\beta_{K^\ast})\,e_1(p_{K^\ast
K^\ast},\beta_{\phi_1} ,\beta_{K^\ast},\beta_{K^\ast}) \right.
\nn\\
&&\left.+c_1^{\phi_\Delta}c_2^{\phi_\Delta} f_{18}(p_{K^\ast
K^\ast},\beta_{\phi_1} ,\beta_{K^\ast}, \beta_{\phi_\Delta})\,
e_2(p_{K^\ast K^\ast},\beta_{\phi_1} ,\beta_{K^\ast},\beta_{K^\ast},
\beta_{\phi_\Delta})
 \right.
\nn\\
&&\left. +c_1^{\omega_\Delta}c_2^{\omega_\Delta} f_{18}(p_{K^\ast
K^\ast},\beta_{\phi_1} ,\beta_{K^\ast}, \beta_{\omega_\Delta})\,
e_2(p_{K^\ast K^\ast},\beta_{\phi_1} ,\beta_{K^\ast},\beta_{K^\ast},
\beta_{\omega_\Delta}) \frac{}{}\right\}
\label{hfic3po-phi1.a}
\end{eqnarray}
\begin{eqnarray}
{\cal C}^{\phi_1 K^\ast K^\ast}_{32}&=&
(c_1^{\phi_1}-c_2^{\phi_1})\left\{\frac{}{}f_{19}(p_{K^\ast
K^\ast},\beta_{\phi_1} ,\beta_{K^\ast})\,e_1(p_{K^\ast
K^\ast},\beta_{\phi_1} ,\beta_{K^\ast},\beta_{K^\ast}) \right.
\nn\\
&&\left.+c_1^{\phi_\Delta}c_2^{\phi_\Delta} f_{20}(p_{K^\ast
K^\ast},\beta_{\phi_1} ,\beta_{K^\ast}, \beta_{\phi_\Delta})\,
e_2(p_{K^\ast K^\ast},\beta_{\phi_1} ,\beta_{K^\ast},\beta_{K^\ast},
\beta_{\phi_\Delta})
 \right.
\nn\\
&&\left. +c_1^{\omega_\Delta}c_2^{\omega_\Delta} f_{20}(p_{K^\ast
K^\ast},\beta_{\phi_1} ,\beta_{K^\ast}, \beta_{\omega_\Delta})\,
e_2(p_{K^\ast K^\ast},\beta_{\phi_1} ,\beta_{K^\ast},\beta_{K^\ast},
\beta_{\omega_\Delta}) \frac{}{}\right\}
\label{hfic3po-phi1.b}
\end{eqnarray}
\begin{eqnarray}
f_{17}(p,\beta_A,\beta_B)&=&\frac{16 \sqrt{2}\, \beta_A^{7/2} p
\left(-5 \left(2 \beta_A^2
\beta_B^2+\beta_B^4\right)+\left(\beta_A^2 +\beta_B^2\right)
p^2\right)}{15 \sqrt{5}\left(2 \beta_A^2+\beta_B^2\right)^{9/2}}
\label{hfic3po-f17}
\end{eqnarray}
\begin{eqnarray}
f_{18}(p,\beta_A ,\beta_B,\beta)&=& \frac{64 \sqrt{2} \beta_A^{7/2}
\beta_B^3 \left(\beta^2+\beta_B^2\right) p \left(5 \left(2 b^2
\beta_A^2 \beta_B^4+2 \left(\beta^2+\beta_A^2\right)
\beta_B^6+\beta_B^8\right) -\left(\beta^2+\beta_B^2\right)^2
\left(\beta_A^2+\beta_B^2\right) p^2\right)} {45\sqrt{5} \left(2 b^2
\beta_A^2+2 \left(\beta^2+\beta_A^2\right)
\beta_B^2+\beta_B^4\right)^{9/2}}\nn\\
\label{hfic3po-f18}
\end{eqnarray}
\begin{eqnarray}
f_{19}(p,\beta_A,\beta_B)&=&-\frac{32 \beta_A^{7/2} \left(\beta_B^2
+\beta_A^2\right) p^3}{5 \sqrt{35} \left(\beta_B^2+2
\beta_A^2\right)^{9/2}}
\label{hfic3po-f19}
\end{eqnarray}
\begin{eqnarray}
f_{20}(p,\beta_A ,\beta_B,\beta)&=& \frac{128 \beta_B^3
\beta_A^{7/2} \left(\beta_B^2+\beta_A^2\right)
\left(\beta_B^2+\beta^2\right)^3 p^3}{15 \sqrt{35} \left(\beta_B^4+2
\beta_A^2 \beta^2+2 \beta_B^2
\left(\beta_A^2+\beta^2\right)\right)^{9/2}}
\label{hfic3po-f20}
\end{eqnarray}






$\rm{\underline{\phi_2(1850)\rightarrow K^+ K^-}:}$

\begin{eqnarray}
{\cal C}^{\phi_2 K K}_{30}&=&\frac{5\sqrt{2}}{12}
(c_1^{\phi_2}-c_2^{\phi_2})\left\{\frac{}{}c_1^{\phi_\Delta}
c_2^{\phi_\Delta}\, f_{20}(p_{K K},\beta_{\phi_2} ,\beta_{K},
\beta_{\phi_\Delta})\, e_2(p_{K K},\beta_{\phi_2}
,\beta_{K},\beta_{K}, \beta_{\phi_\Delta}) \right.
\nn\\
&&\left. +c_1^{\omega_\Delta} c_2^{\omega_\Delta}\, f_{20}(p_{K
K},\beta_{\phi_2} ,\beta_{K}, \beta_{\omega_\Delta})\, e_2(p_{K
K},\beta_{\phi_2} ,\beta_{K},\beta_{K},
\beta_{\omega_\Delta})\frac{}{}\right\} \label{hfic3po-phi2-1}
\end{eqnarray}

$\rm{\underline{\phi_2(1850)\rightarrow K^+ K^{\ast-}}:}$

\begin{eqnarray}
{\cal C}^{\phi_2 K K^\ast}_{11}&=&
(c_1^{\phi_2}+c_2^{\phi_2})\left\{\frac{}{}f_{21}(p_{K
K^\ast},\beta_{\phi_2} ,\beta_{K},\beta_{K^\ast})\, e_1(p_{K
K^\ast},\beta_{\phi_2} ,\beta_{K},\beta_{K^\ast}) \right.
\nn\\
&&\left.  +c_1^{\phi_\Delta} c_2^{\phi_\Delta} f_{22}(p_{K
K^\ast},\beta_{\phi_2} ,\beta_{K},\beta_{K^\ast},
\beta_{\phi_\Delta})\,  e_2(p_{K K^\ast},\beta_{\phi_2}
,\beta_{K},\beta_{K^\ast}, \beta_{\phi_\Delta})
 \right.
\nn\\
&&\left.  +c_1^{\omega_\Delta} c_2^{\omega_\Delta} f_{22}(p_{K
K^\ast},\beta_{\phi_2} ,\beta_{K},\beta_{K^\ast},
\beta_{\omega_\Delta})\,  e_2(p_{K K^\ast},\beta_{\phi_2}
,\beta_{K},\beta_{K^\ast}, \beta_{\omega_\Delta})\frac{}{} \right\}
\label{hfic3po-phi2-2.a}
\end{eqnarray}
\begin{eqnarray}
{\cal C}^{\phi_2 K K^\ast}_{31}&=&
(c_1^{\phi_2}+c_2^{\phi_2})\left\{\frac{}{}f_{23}(p_{K
K^\ast},\beta_{\phi_2} ,\beta_{K},\beta_{K^\ast})\, e_1(p_{K
K^\ast},\beta_{\phi_2} ,\beta_{K},\beta_{K^\ast}) \right.
\nn\\
&&\left.  +c_1^{\phi_\Delta} c_2^{\phi_\Delta} f_{24}(p_{K
K^\ast},\beta_{\phi_2} ,\beta_{K},\beta_{K^\ast},
\beta_{\phi_\Delta})\,  e_2(p_{K K^\ast},\beta_{\phi_2}
,\beta_{K},\beta_{K^\ast}, \beta_{\phi_\Delta})
 \right.
\nn\\
&&\left.  +c_1^{\omega_\Delta} c_2^{\omega_\Delta} f_{24}(p_{K
K^\ast},\beta_{\phi_2} ,\beta_{K},\beta_{K^\ast},
\beta_{\omega_\Delta})\,  e_2(p_{K K^\ast},\beta_{\phi_2}
,\beta_{K},\beta_{K^\ast}, \beta_{\omega_\Delta})\frac{}{} \right\}
\label{hfic3po-phi2-2.b}
\end{eqnarray}
\begin{eqnarray}
f_{21}(p,\beta_A ,\beta_B, \beta_C)&=&\frac{2 \beta_B^{3/2}
\beta_C^{3/2} \left(\beta_B^2+\beta_C^2\right) \beta_A^{7/2} p} {5
\left(\beta_C^2 \beta_A^2+\beta_B^2 \left(\beta_C^2+\beta_A^2
\right)\right)^{9/2}}\left[20 \left(\beta_B^2 \beta_C^4 \beta_A^2
+\beta_B^4 \beta_C^2
\left(\beta_C^2+\beta_A^2\right)\right)-\left(\beta_B^2
+\beta_C^2\right)
 \right.
\nn\\
&&\left. \times\left(2 \beta_B^2
\beta_C^2+\left(\beta_B^2+\beta_C^2\right) \beta_A^2\right)
p^2\right]
\label{hfic3po-f21}
\end{eqnarray}
\begin{eqnarray}
f_{22}(p,\beta_A ,\beta_B, \beta_C,\beta)&=&-\frac{4 \beta_B^{3/2}
\beta_C^{3/2} \beta_A^{7/2} \left(\beta_B^2+\beta_C^2+2
\beta^2\right) p}{45 \left(\beta_C^2 \left(\beta_B^2+2
\beta^2\right)+\beta_A^2 \left(\beta_B^2+\beta_C^2+2
\beta^2\right)\right)^{9/2}}\left[20 \beta_B^2 \beta_C^2
\left(\beta_C^2 \left(\beta_B^2+2 \beta^2\right)+\beta_A^2
\right.\right.
\nn\\
&&\left.\left.\times \left(\beta_B^2+\beta_C^2 +2
\beta^2\right)\right)-\left(\beta_B^2+\beta_C^2+2 \beta^2\right)
\left(2 \beta_C^2 \left(\beta_B^2+\beta^2\right)+\beta_A^2
\left(\beta_B^2+\beta_C^2 +2 \beta^2\right)\right) p^2\right]\nn\\
\label{hfic3po-f22}
\end{eqnarray}
\begin{eqnarray}
f_{23}(p,\beta_A ,\beta_B, \beta_C)&=&-\frac{2 \sqrt{\frac{2}{7}}
\beta_B^{3/2} \beta_C^{3/2} \left(\beta_B^2+\beta_C^2\right)^2
\beta_A^{7/2} \left(2 \beta_B^2 \beta_C^2
+\left(\beta_B^2+\beta_C^2\right) \beta_A^2\right) p^3}{15
\left(\beta_C^2 \beta_A^2+\beta_B^2
\left(\beta_C^2+\beta_A^2\right)\right)^{9/2}}
\label{hfic3po-f23}
\end{eqnarray}
\begin{eqnarray}
f_{24}(p,\beta_A
,\beta_B, \beta_C,\beta)&=& -\frac{4 \sqrt{\frac{2}{7}}
\beta_B^{3/2} \beta_C^{3/2} \beta_A^{7/2}
\left(\beta_B^2+\beta_C^2+2 \beta^2\right)^2 \left(2 \beta_C^2
\left(\beta_B^2+\beta^2\right)+\beta_A^2 \left(\beta_B^2+\beta_C^2+2
\beta^2\right)\right) p^3}{15 \left(\beta_C^2 \left(\beta_B^2+2
\beta^2\right)+\beta_A^2 \left(\beta_B^2+\beta_C^2+2
\beta^2\right)\right)^{9/2}}\nn\\
\label{hfic3po-f24}
\end{eqnarray}

$\rm{\underline{\phi_2(1850)\rightarrow K^{\ast+} K^{\ast-}}:}$

\begin{eqnarray}
{\cal C}^{\phi_2 K^\ast K^\ast}_{12}&=&
-\sqrt{10}\,(c_1^{\phi_2}-c_2^{\phi_2})\left\{\frac{}{}f_{17}(p_{K^\ast
K^\ast},\beta_{\phi_2} ,\beta_{K^\ast})\, e_1(p_{K^\ast
K^\ast},\beta_{\phi_2},\beta_{K^\ast},\beta_{K^\ast}) \right.
\nn\\
&&\left.  +c_1^{\phi_\Delta} c_2^{\phi_\Delta} f_{18}(p_{K^\ast
K^\ast},\beta_{\phi_2},\beta_{K^\ast},\beta_{\phi_\Delta})\,
e_2(p_{K^\ast K^\ast},\beta_{\phi_2},\beta_{K^\ast},\beta_{K^\ast},
\beta_{\phi_\Delta}) \right.
\nn\\
&&\left.  +c_1^{\omega_\Delta} c_2^{\omega_\Delta} f_{18}(p_{K^\ast
K^\ast},\beta_{\phi_2},\beta_{K^\ast},\beta_{\omega_\Delta})\,
e_2(p_{K^\ast K^\ast},\beta_{\phi_2},\beta_{K^\ast},\beta_{K^\ast},
\beta_{\omega_\Delta})\frac{}{}\right\}\nn\\
\label{hfic3po-phi2-3.a}
\end{eqnarray}
\begin{eqnarray}
{\cal C}^{\phi_2 K^\ast K^\ast}_{32}&=&
-\sqrt{\frac{5}{3}}\,(c_1^{\phi_2}-c_2^{\phi_2})\left\{\frac{}{}f_{19}(p_{K^\ast
K^\ast},\beta_{\phi_2} ,\beta_{K^\ast})\, e_1(p_{K^\ast
K^\ast},\beta_{\phi_2},\beta_{K^\ast},\beta_{K^\ast}) \right.
\nn\\
&&\left.  +c_1^{\phi_\Delta} c_2^{\phi_\Delta} f_{20}(p_{K^\ast
K^\ast},\beta_{\phi_2},\beta_{K^\ast},\beta_{\phi_\Delta})\,
e_2(p_{K^\ast K^\ast},\beta_{\phi_2},\beta_{K^\ast},\beta_{K^\ast},
\beta_{\phi_\Delta}) \right.
\nn\\
&&\left.  +c_1^{\omega_\Delta} c_2^{\omega_\Delta} f_{20}(p_{K^\ast
K^\ast},\beta_{\phi_2},\beta_{K^\ast},\beta_{\omega_\Delta})\,
e_2(p_{K^\ast K^\ast},\beta_{\phi_2},\beta_{K^\ast},\beta_{K^\ast},
\beta_{\omega_\Delta})\frac{}{}\right\}\nn\\
\label{hfic3po-phi2-3.b}
\end{eqnarray}



$\rm{\underline{\phi_2(1850)\rightarrow \eta \phi}:}$

\begin{eqnarray}
{\cal C}^{\phi_2 \eta \phi}_{11}&=& -2\left\{\frac{}{}(2 c_1^{\eta}
c_1^{\phi} c_1^{\phi_2}+c_2^{\eta} c_2^{\phi}
c_2^{\phi_2})\,f_{21}(p_{\eta\phi},\beta_{\phi_2}
,\beta_{\eta},\beta_{\phi}) \,e_1(p_{\eta\phi},\beta_{\phi_2}
,\beta_{\eta},\beta_{\phi}) \right.
\nn\\
&&\left. +\left(2 c_1^{\eta} c_1^{\phi} c_1^{\phi_2}
(c_1^{\phi_\Delta})^2+c_2^{\eta} c_2^{\phi} c_2^{\phi_2}
(c_2^{\phi_\Delta})^2\right)f_{22}(p_{\eta\phi},\beta_{\phi_2}
,\beta_{\eta},\beta_{\phi}, \beta_{\phi_\Delta})\,
e_2(p_{\eta\phi},\beta_{\phi_2} ,\beta_{\eta},\beta_{\phi},
\beta_{\phi_\Delta})\right.
\nn\\
&&\left. +\left(2 c_1^{\eta} c_1^{\phi} c_1^{\phi_2}
(c_1^{\omega_\Delta})^2+c_2^{\eta} c_2^{\phi} c_2^{\phi_2}
(c_2^{\omega_\Delta})^2\right)f_{22}(p_{\eta\phi},\beta_{\phi_2}
,\beta_{\eta},\beta_{\phi}, \beta_{\omega_\Delta})\,
e_2(p_{\eta\phi},\beta_{\phi_2} ,\beta_{\eta},\beta_{\phi},
\beta_{\omega_\Delta})\right\} \label{hfic3po-phi2-5.a}
\end{eqnarray}
\begin{eqnarray}
{\cal C}^{\phi_2 \eta \phi}_{31}&=& -2\left\{\frac{}{}(2 c_1^{\eta}
c_1^{\phi} c_1^{\phi_2}+c_2^{\eta} c_2^{\phi}
c_2^{\phi_2})\,f_{23}(p_{\eta\phi},\beta_{\phi_2}
,\beta_{\eta},\beta_{\phi}) \,e_1(p_{\eta\phi},\beta_{\phi_2}
,\beta_{\eta},\beta_{\phi}) \right.
\nn\\
&&\left. +\left(2 c_1^{\eta} c_1^{\phi} c_1^{\phi_2}
(c_1^{\phi_\Delta})^2+c_2^{\eta} c_2^{\phi} c_2^{\phi_2}
(c_2^{\phi_\Delta})^2\right)f_{24}(p_{\eta\phi},\beta_{\phi_2}
,\beta_{\eta},\beta_{\phi}, \beta_{\phi_\Delta})\,
e_2(p_{\eta\phi},\beta_{\phi_2} ,\beta_{\eta},\beta_{\phi},
\beta_{\phi_\Delta})\right.
\nn\\
&&\left. +\left(2 c_1^{\eta} c_1^{\phi} c_1^{\phi_2}
(c_1^{\omega_\Delta})^2+c_2^{\eta} c_2^{\phi} c_2^{\phi_2}
(c_2^{\omega_\Delta})^2\right)f_{24}(p_{\eta\phi},\beta_{\phi_2}
,\beta_{\eta},\beta_{\phi}, \beta_{\omega_\Delta})\,
e_2(p_{\eta\phi},\beta_{\phi_2} ,\beta_{\eta},\beta_{\phi},
\beta_{\omega_\Delta}\frac{}{})\right\} \label{hfic3po-phi2-5.b}
\end{eqnarray}

$\rm{\underline{\phi_3(1850)\rightarrow K^+ K^-}:}$

\begin{eqnarray}
{\cal C}^{\phi_3 K K}_{30}&=&-\frac{5}{4}\,
(c_1^{\phi_3}-c_2^{\phi_3})\left\{\frac{}{}2\,f_{19}(p_{KK},\beta_{\phi_3}
,\beta_{K})\,e_1(p_{KK},\beta_{\phi_3} ,\beta_{K},\beta_{K}) \right.
\nn\\
&&\left. +c_1^{\eta_\Delta} c_2^{\eta_\Delta}
\,f_{20}(p_{KK},\beta_{\phi_3} ,\beta_{K}, \beta_{\eta_\Delta})\,
e_2(p_{KK},\beta_{\phi_3} ,\beta_{K},\beta_{K}, \beta_{\eta_\Delta})
 \right.
\nn\\
&&\left. +c_1^{\eta_\Delta^\prime} c_2^{\eta_\Delta^\prime}
\,f_{20}(p_{KK},\beta_{\phi_3} ,\beta_{K},
\beta_{\eta_\Delta^\prime})\, e_2(p_{KK},\beta_{\phi_3}
,\beta_{K},\beta_{K}, \beta_{\eta_\Delta^\prime}) \right\}
\label{hfic3po-phi3-1}
\end{eqnarray}

$\rm{\underline{\phi_3(1850)\rightarrow K^+ K^{\ast-}}:}$

\begin{eqnarray}
{\cal C}^{\phi_3 K K^\ast}_{31}&=&-\sqrt{\frac{5}{6}}
(c_1^{\phi_3}+c_2^{\phi_3})\left\{\frac{}{}3\,
f_{23}(p_{KK^\ast},\beta_{\phi_3} ,\beta_K,\beta_{K^\ast})\,
e_1(p_{KK^\ast},\beta_{\phi_3} ,\beta_K,\beta_{K^\ast})
 \right.
\nn\\
&&\left.
 +c_1^{\phi_\Delta} c_2^{\phi_\Delta}
\,f_{24}(p_{KK^\ast},\beta_{\phi_3} ,\beta_K,\beta_{K^\ast},
\beta_{\phi_\Delta})\,e_2(p_{KK^\ast},\beta_{\phi_3}
,\beta_K,\beta_{K^\ast}, \beta_{\phi_\Delta})
 \right.
\nn\\
&&\left.
 +c_1^{\omega_\Delta} c_2^{\omega_\Delta}
\,f_{24}(p_{KK^\ast},\beta_{\phi_3} ,\beta_K,\beta_{K^\ast},
\beta_{\omega_\Delta})\,e_2(p_{KK^\ast},\beta_{\phi_3}
,\beta_K,\beta_{K^\ast}, \beta_{\omega_\Delta})\frac{}{}\right\}
\label{hfic3po-phi3-2}
\end{eqnarray}

$\rm{\underline{\phi_3(1850)\rightarrow K^{\ast+} K^{\ast-}}:}$

\begin{eqnarray}
{\cal C}^{\phi_3 K^\ast K^\ast}_{12}&=& 6 \sqrt{\frac{5}{3}}
(c_1^{\phi_3}-c_2^{\phi_3})\left\{\frac{}{}f_{17}(p_{K^\ast
K^\ast},\beta_{\phi_3} ,\beta_{K^\ast})\, e_1(p_{K^\ast
K^\ast},\beta_{\phi_3} ,\beta_{K^\ast},\beta_{K^\ast})
 \right.
\nn\\
&&\left.  + c_1^{\phi_\Delta} c_2^{\phi_\Delta}\, f_{18}(p_{K^\ast
K^\ast},\beta_{\phi_3} ,\beta_{K^\ast}, \beta_{\phi_\Delta})\,
e_2(p_{K^\ast K^\ast},\beta_{\phi_3} ,\beta_{K^\ast},\beta_{K^\ast},
\beta_{\phi_\Delta})
 \right.
\nn\\
&&\left.  + c_1^{\omega_\Delta} c_2^{\omega_\Delta}\,
f_{18}(p_{K^\ast K^\ast},\beta_{\phi_3} ,\beta_{K^\ast},
\beta_{\omega_\Delta})\, e_2(p_{K^\ast K^\ast},\beta_{\phi_3}
,\beta_{K^\ast},\beta_{K^\ast}, \beta_{\omega_\Delta})\frac{}{}
\right\}
 \label{hfic3po-phi3-3.a}
\end{eqnarray}
\begin{eqnarray}
{\cal C}^{\phi_3 K^\ast K^\ast}_{32}&=&\sqrt{\frac{5}{3}}
(c_1^{\phi_3}-c_2^{\phi_3})\left\{\frac{}{}f_{19}(p_{K^\ast
K^\ast},\beta_{\phi_3} ,\beta_{K^\ast})\, e_1(p_{K^\ast
K^\ast},\beta_{\phi_3} ,\beta_{K^\ast},\beta_{K^\ast})
 \right.
\nn\\
&&\left.  + c_1^{\phi_\Delta} c_2^{\phi_\Delta}\, f_{20}(p_{K^\ast
K^\ast},\beta_{\phi_3} ,\beta_{K^\ast}, \beta_{\phi_\Delta})\,
e_2(p_{K^\ast K^\ast},\beta_{\phi_3} ,\beta_{K^\ast},\beta_{K^\ast},
\beta_{\phi_\Delta})
 \right.
\nn\\
&&\left.  + c_1^{\omega_\Delta} c_2^{\omega_\Delta}\,
f_{20}(p_{K^\ast K^\ast},\beta_{\phi_3} ,\beta_{K^\ast},
\beta_{\omega_\Delta})\, e_2(p_{K^\ast K^\ast},\beta_{\phi_3}
,\beta_{K^\ast},\beta_{K^\ast}, \beta_{\omega_\Delta})
\frac{}{}\right\}
 \label{hfic3po-phi3-3.b}
\end{eqnarray}

$\rm{\underline{\phi_3(1850)\rightarrow K^+ K_1^-(1270)}:}$

\begin{eqnarray}
{\cal C}^{\phi_3 KK_1(1270)}_{21}&=& (c_1^{\phi_{2050}} +
c_2^{\phi_3})\left\{\frac{}{}f_{25}(p_{KK_1},\beta_{\phi_3}
,\beta_{K},\beta_{K_1})\, e_1(p_{KK_1},\beta_{\phi_3},
\beta_K,\beta_{K_1}) \right.
\nn\\
&&\left. + c_1^{\eta_\Delta}c_2^{\eta_\Delta} f_{26}(p_{K
K_1},\beta_{\phi_3}, \beta_{K},\beta_{K_1},\beta_{\eta_\Delta})\,
e_2(p_{K K_1},\beta_{\phi_3},
\beta_{K},\beta_{K_1},\beta_{\eta_\Delta}) \right.
\nn\\
&&\left.+ c_1^{\eta_\Delta^\prime}
c_2^{\eta_\Delta^\prime}f_{26}(p_{K K_1},\beta_{\phi_3},
\beta_{K},\beta_{K_1},\beta_{\eta^\prime_\Delta})\, e_2(p_{K
K_1},\beta_{\phi_3},
\beta_{K},\beta_{K_1},\beta_{\eta^\prime_\Delta})\frac{}{}\right\}\cos\theta
\nn\\
&& + (c_1^{\phi_3} -
c_2^{\phi_3})\left\{\frac{}{}f_{27}(p_{KK_1},\beta_{\phi_3}
,\beta_{K},\beta_{K_1})\, e_1(p_{KK_1},\beta_{\phi_3},
\beta_K,\beta_{K_1})\right.
\nn\\
&&\left. + c_1^{\phi_\Delta} c_2^{\phi_\Delta}f_{28}(p_{K
K_1},\beta_{\phi_3}, \beta_{K},\beta_{K_1},\beta_{\phi_\Delta})\,
e_2(p_{K K_1},\beta_{\phi_3},
\beta_{K},\beta_{K_1},\beta_{\phi_\Delta})\right.
\nn\\
&&\left. + c_1^{\omega_\Delta} c_2^{\omega_\Delta}f_{28}(p_{K
K_1},\beta_{\phi_3}, \beta_{K},\beta_{K_1},\beta_{\omega_\Delta})\,
e_2(p_{K K_1},\beta_{\phi_3},
\beta_{K},\beta_{K_1},\beta_{\omega_\Delta})\frac{}{}\right\}\sin\theta
\label{hfic3po-phi3-4.a}
\end{eqnarray}
\begin{eqnarray}
 {\cal C}^{\phi_3KK_1(1270)}_{41}&=& (c_1^{\phi_3}
+ c_2^{\phi_3})\left\{\frac{}{}f_{29}(p_{KK_1},\beta_{\phi_3}
,\beta_{K},\beta_{K_1})\, e_1(p_{KK_1},\beta_{\phi_3},
\beta_K,\beta_{K_1}) \right.
\nn\\
&&\left.+ c_1^{\eta_\Delta}c_2^{\eta_\Delta}f_{30}(p_{K
K_1},\beta_{\phi_3}, \beta_{K},\beta_{K_1},\beta_{\eta_\Delta})\,
e_2(p_{K K_1},\beta_{\phi_3},
\beta_{K},\beta_{K_1},\beta_{\eta_\Delta}) \right.
\nn\\
&&\left.+c_1^{\eta_\Delta^\prime}
c_2^{\eta_\Delta^\prime}f_{30}(p_{K K_1},\beta_{\phi_3},
\beta_{K},\beta_{K_1},\beta_{\eta^\prime_\Delta})\, e_2(p_{K
K_1},\beta_{\phi_3},
\beta_{K},\beta_{K_1},\beta_{\eta^\prime_\Delta})
\frac{}{}\right\}\cos\theta
\nn\\
&& + (c_1^{\phi_3} -
c_2^{\phi_3})\sqrt{2}\left\{-\frac{1}{2}\,f_{29}(p_{KK_1},\beta_{\phi_3}
,\beta_{K},\beta_{K_1})\, e_1(p_{KK_1},\beta_{\phi_3},
\beta_K,\beta_{K_1})\right.
\nn\\
&&\left. + c_1^{\phi_\Delta} c_2^{\phi_\Delta}f_{30}(p_{K
K_1},\beta_{\phi_3}, \beta_{K},\beta_{K_1},\beta_{\phi_\Delta})\,
e_2(p_{K K_1},\beta_{\phi_3},
\beta_{K},\beta_{K_1},\beta_{\phi_\Delta})\right.
\nn\\
&&\left. + c_1^{\omega_\Delta} c_2^{\omega_\Delta}f_{30}(p_{K
K_1},\beta_{\phi_3}, \beta_{K},\beta_{K_1},\beta_{\omega_\Delta})\,
e_2(p_{K K_1},\beta_{\phi_3},
\beta_{K},\beta_{K_1},\beta_{\omega_\Delta}\frac{}{})\right\}\sin\theta
\label{hfic3po-phi3-4.b}
\end{eqnarray}
\begin{eqnarray}
f_{25}(p,\beta_A ,\beta_B,\beta_C)&=&\frac{\sqrt{2}\, \beta_B^{7/2}
\beta_C^{5/2} \left(\beta_B^2+\beta_C^2\right) \beta_A^{7/2}
p^2}{7\sqrt{15} \left(\beta_C^2 \beta_A^2+\beta_B^2
\left(\beta_C^2+\beta_A^2\right) \right)^{11/2}}\left[28 \left(4
\beta_B^4 \beta_C^4+5 \beta_B^2 \beta_C^2
\left(\beta_B^2+\beta_C^2\right)
\beta_A^2+\left(\beta_B^2+\beta_C^2\right)^2 \beta_A^4\right)
\right.
\nn\\
&&\left.-3 \left(\beta_B^2+\beta_C^2\right) \left(2 \beta_B^2
\beta_C^2+\left(\beta_B^2+\beta_C^2\right) \beta_A^2\right)
p^2\right]
\label{hfic3po-f25}
\end{eqnarray}
\begin{eqnarray}
f_{26}(p,\beta_A ,\beta_B,\beta_C,\beta)&=&-\frac{2 \sqrt{2}\,
\beta_B^{3/2} \beta_C^{5/2} \left(\beta_B^2+\beta_C^2+2
\beta^2\right) \beta_A^{7/2}p^2}{21 \sqrt{15}\left(\beta_C^2
\left(\beta_B^2+2 \beta^2\right) +\left(\beta_B^2+\beta_C^2+2
\beta^2\right) \beta_A^2\right)^{11/2}} \left[28 \left(4 \beta_C^4
\left(\beta_B^2+\beta^2\right) \left(\beta_B^2+2 \beta^2\right)^2
\right.\right.
\nn\\
&&\left.\left.+\beta_C^2 \left(\beta_B^2+\beta_C^2 +2 \beta^2\right)
\left(5 \beta_B^4+18 \beta_B^2 \beta^2+16 \beta^4\right)
\beta_A^2+\left(\beta_B^2+\beta_C^2+2 \beta^2\right)^2
\left(\beta_B^2 +4 \beta^2\right) \beta_A^4\right)\right.
\nn\\
&&\left.-3 \left(\beta_B^2+2 \beta^2\right)
\left(\beta_B^2+\beta_C^2+2 \beta^2\right) \left(2 \beta_C^2
\left(\beta_B^2+\beta^2\right)+\left(\beta_B^2+\beta_C^2+2
\beta^2\right) \beta_A^2\right) p^2\right]
\label{hfic3po-f26}
\end{eqnarray}
\begin{eqnarray}
f_{27}(p,\beta_A ,\beta_B,\beta_C)&=&\frac{4 \beta_B^{7/2}
\beta_C^{5/2} \left(\beta_B^2+\beta_C^2\right) \beta_A^{7/2} p^2}{7
\sqrt{15} \left(\beta_C^2 \beta_A^2+\beta_B^2
\left(\beta_C^2+\beta_A^2\right) \right)^{11/2}}\left[28
\left(\beta_B^2 \beta_C^4 \beta_A^2+\beta_B^4 \beta_C^2
\left(\beta_C^2+\beta_A^2\right)\right)-\left(\beta_B^2+\beta_C^2\right)\right.
\nn\\
&&\left. \left(2 \beta_B^2
\beta_C^2+\left(\beta_B^2+\beta_C^2\right) \beta_A^2\right)
p^2\right]
\label{hfic3po-f27}
\end{eqnarray}
\begin{eqnarray}
f_{28}(p,\beta_A ,\beta_B,\beta_C,\beta)&=&-\frac{2 \beta_B^{3/2}
\beta_C^{5/2} \beta_A^{7/2} \left(\beta_B^2+\beta_C^2+2
\beta^2\right)^2 p^2} {21 \sqrt{15} \left(\beta_C^2
\beta_A^2+\beta_B^2 \left(\beta_C^2+\beta_A^2\right) +2
\left(\beta_C^2+\beta_A^2\right) \beta^2\right)^{11/2}}\left[28\,
\beta_B^2 \beta_A^2 \left(\beta_C^2 \beta_A^2+\beta_B^2
\left(\beta_C^2+\beta_A^2\right) \right.\right.
\nn\\
&&\left.\left.+2 \left(\beta_C^2+\beta_A^2\right)
\beta^2\right)+\left(\beta_B^2+2 \beta^2\right) \left(\beta_C^2
\beta_A^2+\beta_B^2 \left(2 \beta_C^2+\beta_A^2\right) +2
\left(\beta_C^2+\beta_A^2\right) \beta^2\right)
p^2\right]
\label{hfic3po-f28}
\end{eqnarray}
\begin{eqnarray}
f_{29}(p,\beta_A ,\beta_B,\beta_C)&=&\frac{\sqrt{2} \beta_B^{7/2}
\beta_C^{5/2} \left(\beta_B^2+\beta_C^2\right)^2 \beta_A^{7/2}
\left(2 \beta_B^2 \beta_C^2+\left(\beta_B^2+\beta_C^2\right)
\beta_A^2\right) p^4 } {21\sqrt{5} \left(\beta_C^2
\beta_A^2+\beta_B^2 \left(\beta_C^2
+\beta_A^2\right)\right)^{11/2}}
\label{hfic3po-f29}
\end{eqnarray}
\begin{eqnarray}
f_{30}(p,\beta_A ,\beta_B,\beta_C,\beta)&=&-\frac{2 \sqrt{2}
\beta_B^{3/2} \beta_C^{5/2} \left(\beta_B^2+2 \beta^2\right)
\left(\beta_B^2+\beta_C^2+2 \beta^2\right)^2
\beta_A^{7/2}p^4}{63\sqrt{5} \left(\beta_C^2 \left(\beta_B^2+2
\beta^2\right)+\left(\beta_B^2+\beta_C^2+2 \beta^2\right)
\beta_A^2\right)^{11/2}}
\label{hfic3po-f30}
\end{eqnarray}

$\rm{\underline{\phi_3(1850)\rightarrow \eta \phi}:}$

\begin{eqnarray}
{\cal C}^{\phi_3 \eta \phi}_{31}&=& \sqrt{\frac{10}{3}}\left\{3\,(2
c_1^{\eta} c_1^{\phi} c_1^{\phi_3}+c_2^{\eta} c_2^{\phi}
c_2^{\phi_3})\,f_{23}(p_{\eta\phi},\beta_{\phi_3}
,\beta_\eta,\beta_\phi)\, e_1(p_{\eta\phi},\beta_{\phi_3}
,\beta_\eta,\beta_\phi)
 \right.
\nn\\
&&\left. +\left(2 c_1^{\eta} c_1^{\phi} c_1^{\phi_3}
(c_1^{\phi_\Delta})^2+c_2^{\eta} c_2^{\phi} c_2^{\phi_3}
(c_2^{\phi_\Delta})^2\right) f_{24}(p_{\eta\phi},\beta_{\phi_3}
,\beta_\eta,\beta_\phi, \beta_{\phi_\Delta}) \,
e_2(p_{\eta\phi},\beta_{\phi_3} ,\beta_\eta,\beta_\phi,
\beta_{\phi_\Delta}) \right.
\nn\\
&&\left. +\left(2 c_1^{\eta} c_1^{\phi} c_1^{\phi_3}
(c_1^{\omega_\Delta})^2+c_2^{\eta} c_2^{\phi} c_2^{\phi_3}
(c_2^{\omega_\Delta})^2\right) f_{24}(p_{\eta\phi},\beta_{\phi_3}
,\beta_\eta,\beta_\phi, \beta_{\omega_\Delta})\,
 e_2(p_{\eta\phi},\beta_{\phi_3} ,\beta_\eta,\beta_\phi,
\beta_{\omega_\Delta})\right\}
\label{hfic3po-phi3-5}
\end{eqnarray}

\end{widetext}

\end{document}